\pdfoutput=1 
\documentclass[twocolumn,aps,prc,preprintnumbers]{revtex4}%
\usepackage{amssymb}
\usepackage{amsmath}
\usepackage{amsfonts}
\usepackage{graphicx}
\usepackage{dcolumn}
\usepackage{bm}%
\begin{document}
\title{NUCLEAR REACTIONS}
\author{C.A. BERTULANI}

\address{
Department of Physics, Texas A\&M University, Commerce, TX 75429,
USA}\email{carlos_bertulani@tamu-commmerce.edu}

\vspace{2cm}

\underline{\sl Giovanni Raciti}, in memoriam.

\begin{abstract}
Nuclear reactions generate energy in nuclear reactors, in stars, and are responsible for the existence of all elements heavier than hydrogen in the universe. Nuclear reactions denote reactions between nuclei, and between nuclei and other fundamental particles, such as electrons and photons. A short description of the conservation laws and the definition of basic physical quantities is presented, followed by a more detailed account of specific cases: (a) formation and decay of compound nuclei;  (b)direct reactions; (c) photon and electron scattering; (d) heavy ion collisions; (e) formation of a quark-gluon plasma; (f) thermonuclear reactions; (g) and reactions with radioactive beams.  Whenever necessary, basic equations are introduced to help understand general properties of these reactions.     Published in Wiley Encyclopedia of Physics, ISBN-13: 978-3-527-40691-3 - Wiley-VCH, Berlin, 2009.
\end{abstract}


\maketitle

\tableofcontents

\section{Introduction}

The collision  of two nuclei can give place to a nuclear reaction
and, similarly to a chemical reaction, the final products can be
different from the initial ones. This process happens when a target
is bombarded by particles coming from an accelerator or from a
radioactive substance. It was in the latter way that Rutherford
observed, in 1919, the first nuclear reaction produced in
laboratory,
\begin{equation}\alpha + \ ^ {14}_ {\ 7}\hbox{N} \rightarrow \ ^
{17}_ {\ 8}\hbox{O }+\hbox{p}, \label{ru1}
\end{equation} using $\alpha$-particles from a $^ {214}$Bi
sample.

As in eq. \ref{ru1}, other reactions were induced using
$\alpha$-particles, the only projectile available initially. With
the development of accelerators around 1930, the possibilities
multiplied by changing the energy and mass of the projectile. Today
it is possible to bombard a target with protons of energy greater
than 1 TeV (1 TeV = $10^{12}$ eV $ = 1.602 \times 10^{-7}$ joules),
and beams of particles as heavy as uranium are available for study
of reactions with heavy ions.

Sometimes we can have more than two final products in a reaction, as
in
\begin{eqnarray}\hbox{p }+ \ ^ {14}\hbox{N}
\rightarrow \ ^ {7}\hbox{Be }+ \ 2\alpha, \nonumber \\ \hbox{p }+ \
^ {23}\hbox{Na} \rightarrow \ ^ {22}\hbox{Ne }+\hbox {p}+\hbox {n},
\end{eqnarray}
or just one, as in the capture reaction $\hbox{p }+ \ ^
{27}\hbox{Al} \rightarrow \ ^ {28}\hbox{Si}^*,\label{2} $ where the
asterisk indicates  an excited state, which usually decays emitting
$\gamma$-radiation. Under special circumstances, more than two
reactants is possible. Thus, for example, the reaction $\alpha+ \
\alpha+ \ \alpha \ \rightarrow \ ^ {12}\hbox{C}$ can take place in
the overheated plasma of stellar interiors.

The initial and final products can also be identical. This case
characterizes a process which can be elastic, as in$\hbox{p }+ \ ^
{16}\hbox{O} \rightarrow \ \hbox{p }+ \ ^ {16}\hbox{O},\label{4} $
where there is only transfer of kinetic energy between projectile
and target, or inelastic, as in $\hbox{n }+ \ ^ {16}\hbox{O}
\rightarrow \ \hbox{n }+ \ ^ {16} \hbox{O}^*,$ where part of the
kinetic energy of the system is used in the excitation of $^ {16}$O.

Naturally, nuclear reactions are not limited to nuclei. They could
involve any  type of particle, and also radiation. Thus, the
reactions
\begin{eqnarray}\gamma \ + \ ^ {63}\hbox{Cu} & \rightarrow &
\ ^ {62}\hbox{Ni} \ + \ \hbox{p},\cr \gamma \ + \ ^ {233}\hbox{U} &
\rightarrow & \ ^ {90}\hbox{Rb }+ \ ^ {141}\hbox{Cs }+ \ 2\hbox{n},
\end{eqnarray}
are examples of nuclear processes induced by gamma radiation. In the
first case a $\gamma$-ray  knocks a proton off  $^ {63}$Cu and in
the second it induces nuclear fission of  $^ {233}$U, with the
production of two fragments and emission of two neutrons.

Unlike a chemical reaction, the resulting products of a nuclear
reaction are not determined univocally: starting from two or more
reactants there can exist dozens of final products with an unlimited
number of available quantum states. As an example, the collision of
a deuteron with  $^ {238}$U can give place, among others, to the
following reactions:
\smallskip
\begin{eqnarray}\hbox{d} \ + \ ^ {238}\hbox{U}&\rightarrow
\ ^ {240}\hbox{Np} \ + \ \gamma,\cr & \rightarrow \ ^ {239}\hbox{Np}
\ + \ \hbox{n},\cr &\rightarrow \ \phantom{p}^{239}\hbox{U} \ + \
\hbox{p},\cr &\rightarrow \ \phantom{p}^{237}\hbox{U} \ + \
\hbox{t}.\label{channels}\end{eqnarray} In the first of them the
deuteron is absorbed by the uranium, forming an excited nucleus of
$^ {240}$Np that de-excites by emitting a $\gamma$-ray. The two
following are examples of {\it stripping reactions}, in which a
nucleon is transferred from the projectile to the target. The last
one exemplifies the inverse process: the deuteron captures a neutron
from the target and emerges out as $^3$H (tritium). This is denoted
as a {\it pick-up reaction}. Another possibility would be, in the
first reaction, that $^ {240}$Np fissions instead of emitting a
$\gamma$-ray, contributing with dozens of possible final products
for the reaction.

Each reaction branch, with well defined quantum states of the
participants,  is known as {\it channel}. In \ref{channels}, for the
{\it entrance channel} d + $^ {238}$U, there are four possible {\it
exit channels}. Notice that a different exit channel would be
reached if some of the final products were in an excited state. The
probability that a nuclear reaction takes place through a certain
exit channel depends on the energy of the incident particle and is
measured by the {\it cross section} for that channel.

Nuclear reactions proceed through many possible distinct mechanisms.
For instance, in {\it direct reactions} the projectile and the
target have an interaction of short duration, with possible exchange
of energy or particles between them. Another mechanism involves the
{\it fusion} of the projectile with the target, the available energy
being distributed among all nucleons, forming a highly excited {\it
compound nucleus}. The decay of the compound nucleus leads to the
final products of the reaction.

In high energies collisions, the nuclei fragment and particles that
were not initially present are produced (for instance, pions, kaons,
etc.). The reactions proceed through an intermediate phase in which
the nuclear matter is compressed. At very high energies the quarks
and gluons inside nucleons become ``free'' for a short time forming
the {\it quark-gluon plasma}. The study of high energy reactions
with nuclei is very important for a better understanding of what
happens during spectacular stellar explosions (supernovae) and in
the interior of compact stars, as for instance, neutron stars. The
study of nuclear reactions at high energies allows to obtain
information on the {\it equation of state} of nuclear matter.

\section{Basic principles}
\subsection{Conservation laws} Several conservation laws contribute
to restrict the possible processes in a nuclear reaction.

1-  {\it Baryonic number} - There is no experimental evidence of
processes in which  nucleons are created or destroyed without the
creation or destruction of corresponding antinucleons.  The
application of this principle to low energy reactions is still more
restrictive. Below the threshold for the production of mesons
($\sim140 $ MeV), no process related to the nuclear forces is
capable to transform a proton into a neutron and vice-versa, and
processes governed by the weak force (responsible for the
$\beta$-emission of nuclei) are very slow in relation to the times
involved in nuclear reactions ($\sim 10^{-22} $ to $10^{-16} $s). In
this way, we can speak separately of proton and neutron
conservation, which should show up with same amounts in both sides
of a nuclear reaction.

2 - {\it Charge} - This is a general conservation principle in
physics, valid in any circumstance. In purely nuclear  reactions it
is computed making the sum of the atomic numbers, which should be
identical, at both sides of the reaction.

3 - {\it Energy and linear momentum} - These are two of the most
applied principles in the study of the kinematics of reactions.
Through them, angles and velocities are related to the initial
parameters of the problem.

4 - {\it Total angular momentum} - is always a constant of motion.
In the reaction
\begin{equation} ^ {10}\hbox{B }+ \ ^ {4}\hbox{He}
\rightarrow \ ^ {1}\hbox{H }+ \ ^ {13}\hbox{C},\label{10.8}
\end{equation}
$^ {10}$B has $I=3 $ in the ground state, whereas the
$\alpha$-particle has zero angular momentum. If it is captured in an
s-wave ($l_i=0$), the intermediate compound nucleus is in a state
with  $I_{c}=3$. Both final products have intrinsic angular momenta
equal to 1/2. Hence, their sum is 0 or 1. Therefore the relative
angular momentum of the final products will be $l_f $ = 2, 3 or 4.

5 - {\it Parity} - is always conserved in reactions governed by the
nuclear interaction. In the previous example, $^ {10}$B, $^4$He and
the proton have equal parities, while $^ {13}$C has odd parity.
Therefore, if $l_i=0$, we necessarily have $l_f=3$. Thus, the
orbital momentum of the final products of eq. \ref{10.8} is
determined by the joint conservation of the total angular momentum
and of the parity.

6 - {\it Isospin} - This is an approximate conservation law that is
applied to light nuclei, where the effect of the Coulomb force is
small. A nuclear reaction involving these nuclei not only conserves
the $z $-component of the isospin (a consequence of charge and
baryonic number conservation) but also the total isospin {\bf T}.
Reactions that populate excited states not conserving the value of
{\bf T} are strongly inhibited. An example is  the reaction d + $^
{16}$O $\rightarrow\alpha $+ $^ {14}$N, where the excited state $0^+
$, with 2.31 MeV, of $^ {14}$N is about a hundred times more
populated than the $1^+ $ ground state. Conservation of energy,
angular momentum and parity do not impose any prohibition for that
channel, whose low occurrence can only be justified by isospin
conservation: the ground states of the four participant nuclei in
the reaction have all ${\bf T}=0 $ and the state $1^+ $ of $^ {14}$N
has ${\bf T}=1 $.

\subsection{Kinematics} We consider a typical
reaction, in which the projectile a and  the target A gives place to
two products, b and B. This can also be expressed in the notation
that we used so far, $
\hbox{a}+\hbox{A}\rightarrow\hbox{b}+\hbox{B},\label{10.9} $ or even
in a more compact notation, $\hbox{A(a,b)B}. \label{10.10}
$

Often, a and b are light nuclei and A and B, heavy ones; the nucleus
b being emitted at an angle $\theta $ and its energy registered in
the laboratory system. The recoiling nucleus  B has usually a short
range and cannot leave the target. It is convenient to introduce the
{\it Q-value of the reaction} which measures the energy gained (or
lost) due to the difference between the initial and final masses:
\begin{equation}
Q=(m_a+m_A-m_b-m_B)c^2.\label{10.12} \end{equation}

Using energy and momentum conservation in the reaction, one gets:
\begin{eqnarray}
Q&=&E_b\left(1+{m_b\over m_B}\right)-E_a\left(1-{m_a\over
m_B}\right)\nonumber \\
&-& {2\over m_B}\sqrt{m_am_bE_aE_b}\cos\theta.
\end{eqnarray}

\begin{figure}
[t]
\begin{center}
\includegraphics[
natheight=1.5in, natwidth=1.5in, height=3.2in, width=3.2in
]%
{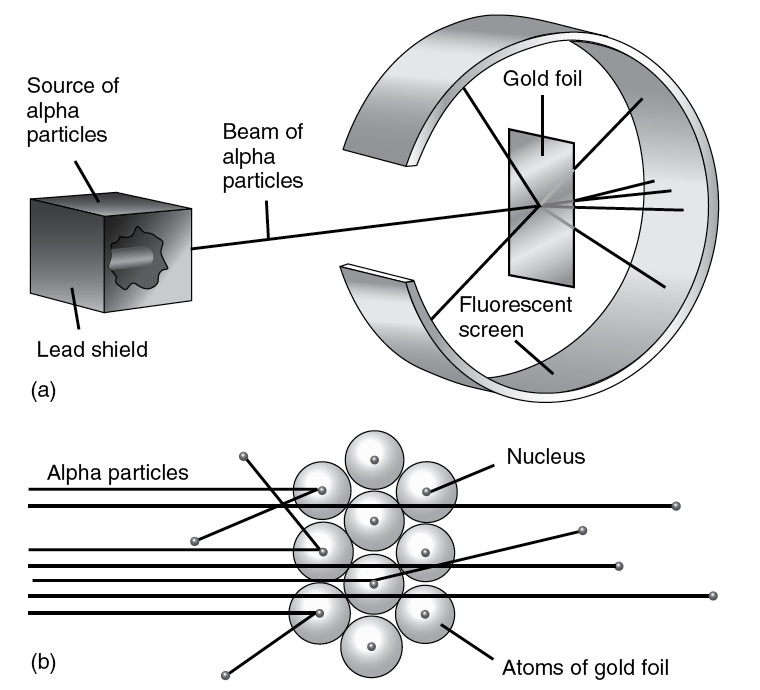}%
\caption{(Top) Rutherford's scattering experiment with
$\alpha$-particles scattering on a gold foil.
(Bottom) Microscopic interpretation of the experimental results.}%
\label{setup}%
\end{center}
\end{figure}

From this relation one obtains that, when $Q $ is negative, an
energy threshold exists for the incident particle, $E_t $, as a
function of the angle $\theta $, below which nuclei b are not
observed in that angle:
\begin{equation}
E_t={-Qm_B(m_B+m_b)\over m_am_b\hbox{cos}^2\theta+(m_B+m_b)
(m_B-m_a)}.\label{10.15} \end{equation}

This example shows the power of conservation laws in the analysis of
nuclear reactions. Similar to this example, very useful relations
can be derived in the relativistic regime, and/or using other
conserved quantities.

\subsection{Cross sections}

Figure \ref{setup} shows schematically a typical scattering
experiment. In fact this is the sketch of E. Rutherford experiment
in 1910 \cite{1Ru11}. Projectiles (here, $\alpha$-particles) from a
source pass through a collimator and collide with a target (gold
foil). Some projectiles are scattered by the target and reach the
detector (here, a fluorescent screen). Rutherford expected them to
go straight through the target, with perhaps minor deflections. Most
did go straight through, but to his surprise some particles bounced
straight back!  Rutherford hypothesized that the positive alpha
particles had hit a heavy mass of positive particles, which he
called {\it the nucleus}.

\begin{figure}
[t]
\begin{center}
\includegraphics[
natheight=1.5in, natwidth=1.5in, height=1.6in, width=3.2in
]%
{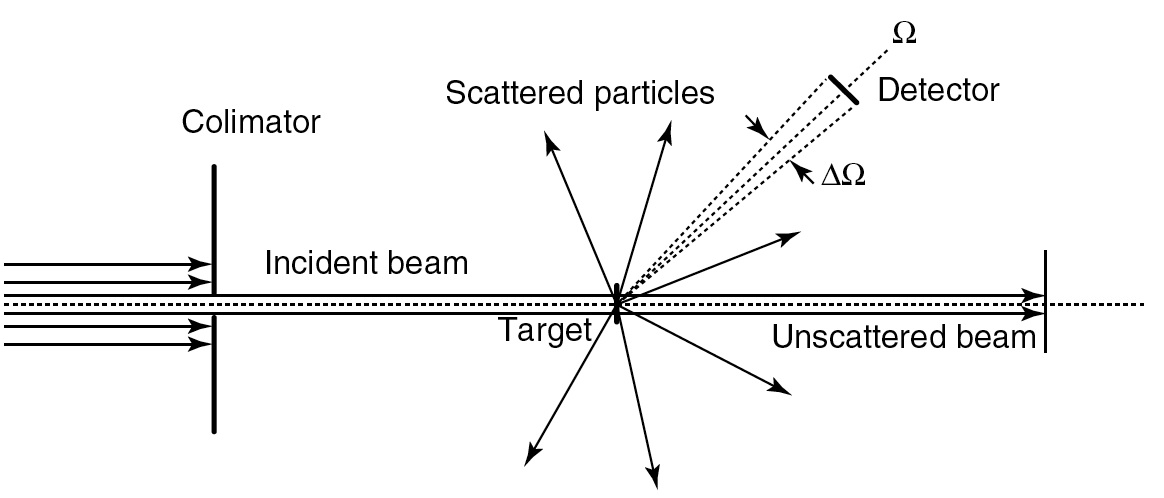}%
\caption{Schematic representation of a scattering experiment.}
\label{exper}
\end{center}
\end{figure}

Let us consider an experiment as that represented in figure
\ref{exper}, measuring the count rate of events leading to the
population of channel-$\alpha$, $N_{\alpha}(\Omega,\Delta\Omega)$.
Assuming that interactions between beam particles can be neglected,
the count rate $N_{\alpha}(\Omega,\Delta\Omega)$ should be
proportional to the incident flux $J$, to the fraction of solid
angle $\Delta \Omega$ where the particles are scattered to, and to
the number of target particles per unit volume. Hence,
$N_{\alpha}(\Omega,\Delta\Omega)$ can then be written as
$N_{\alpha}(\Omega,\Delta\Omega)=\left(  \Delta\Omega\cdot
n\cdot\,J\right) \, d\sigma_{\alpha}(\Omega)/d\Omega$. The constant
of proportionality,
\begin{equation}
\frac{d\sigma_{\alpha}(\Omega)}{d\Omega}=\frac{N_{\alpha}(\Omega,\Delta
\Omega)}{\Delta\Omega\cdot n\cdot J}\,\,,\label{defxsec}%
\end{equation}
is called the {\it differential cross section} for channel-$\alpha$.
This quantity is very useful, since it does not depend on
experimental details (detector size, incident flux, target
thickness). It depends exclusively on the physics of the projectile
and target particles.

Frequently, one is interested in the angle integrated {\it total
cross section} given by
\begin{equation}
\sigma_{\alpha}=\int d\Omega\,\,\left[
\frac{d\sigma_{\alpha}(\Omega)}{d\Omega
}\right]  \,\,.\label{elxsec1}%
\end{equation}

In nuclear physics cross sections are usually measured in units of
{\it barns}. 1 barn corresponds to an area of a circle of diameter
approximately equal to 8 fm = 8 $\times 10^{-15}$ m. For the
comparison between theory and experiment, it is necessary to have
cross sections in the same reference frame. Of course, the measured
cross section is obtained in the laboratory frame, where the target
is at rest. From the theoretical point of view, however, it is
important to take advantage of the translational invariance of the
projectile-target Hamiltonian, introducing the center of mass frame
(CM). The transformation is done using the laws of conservation of
energy and momentum.

The Rutherford experiment has a simple classical description in
which one assumes that the $\alpha$-particles follow hyperbolic
trajectories in the Coulomb field of the heavy target. The
scattering at an angle $\theta$ can be easily calculated and with
that one is able to calculate the Coulomb, or Rutherford cross
section. The result is
\begin{equation}{d\sigma_R\over
d\Omega}=\Bigl({Z_PZ_Te^2\over4E}\Bigr)^2
\hbox{cosec}^4\Bigl({\theta\over2}\Bigr)\
,\label{cou2}\end{equation} where $Z_P$ ($Z_T$) is the projectile
(target) charge number and $E$ is the CM energy.

\subsection{Elastic scattering}

When a beam of particles, represented in quantum mechanics by a
plane wave, hits a nucleus the wave function is modified by the
presence of a scattering potential  $V(r) $, responsible for the
appearance of a phase in the outgoing part of the wave. Elastic
scattering is just one of the channels for the which the reaction
can proceed and is known as {\it elastic channel}. Inelastic
scattering and all the other channels are grouped in the {\it
reaction channel}.

The occurrence of a nuclear reaction through a given reaction
channel leads to a modification of the outgoing part of the wave
function  not only by  a phase factor, but also by changing its
magnitude, indicating that there is a loss of particles in the
elastic channel. For a projectile with momentum $p=\hbar k$, this
can be expressed by
\begin{equation}
\Psi\sim{1\over2i}\sum^\infty_{l=0}(2l+1)i^lP_l(\hbox{cos}\theta){\eta_le^{i(kr-l\pi/2)}
-e^{-i(kr-l\pi/2)} \over kr},\label{Psiscatt} \end{equation} where
the complex coefficient $\eta_l $ is the factor mentioned above. If
$\eta=1$, the sum in eq. \ref{Psiscatt} can be done analytically,
leading to $\Psi \sim \exp(i{\bf k.r})$, i.e. a plane wave. But if
$\eta_l=\exp[i\delta_l]$, with $\delta_l$ real, the incoming and
outgoing waves have the same magnitude, i.e. the scattering is
elastic.

The sum in eq. \ref{Psiscatt} is know as {\it partial wave
expansion} of the scattering wave. The label $l=0, 1, 2 ,\cdots = $
(s, p, d, $\cdots$ waves) denotes the contribution of a particular
angular momentum (in units of $\hbar$) to the total wavefunction.
Classically the angular momentum of an incident particle is given by
$l=kR$, where $R$ is known as {\it impact parameter} which is the
perpendicular distance to the target if the projectile were
undeflected. In quantum mechanics $l$ is not continuous, varying in
steps of one.

We can rewrite eq. \ref{Psiscatt} as a sum of a plane wave and a
scattering outgoing wave in the form $\Psi \sim \exp(i{\bf
k.r})+f(\theta) e^{ikr}/r$, where $f(\theta)$ accounts for the
distortion of the outgoing wave at the scattering angle $\theta$. It
is known as the {\it scattering amplitude}:
\begin{equation}
f(\theta)={1\over2ik}\sum_{
l=0}^\infty(2l+1)(\eta_l-1)P_l(\hbox{cos}\theta). \label{ftheta}
\end{equation}
One can now calculate the cross section by counting the number of
particles that are scattered through angle $\theta$. This can be
done by calculating the particle current associated with the
wavefunction $\Psi$. One gets the differential scattering cross
section
\begin{equation}
{d\sigma_e\over d\Omega}=\vert
f(\theta)\vert^2={1\over4k^2}\biggl\vert\sum^\infty_{l=0}(2l+1)(1-
\eta_l)P_l(\hbox{cos}\theta)\biggr\vert^2.\label{sige}
\end{equation} The total scattering cross section, eq. \ref{elxsec1} becomes
\begin{equation}
\sigma_e=\pi
\bar{\lambda}^2\sum^\infty_{l=0}(2l+1)\vert1-\eta_l\vert^2,\label{sige2}
\end{equation}
with $\bar{\lambda}=\lambda/2\pi=1/k $.

\subsection{Reaction cross sections}

To calculate the reaction cross section it is necessary to compute
the number of particles that disappear from the elastic channel,
what is measured by the flux of the current of probability vector
through a spherical surface of large radius centered at the target,
calculated with the total wave function of eq. \ref{Psiscatt}. One
finds
\begin{equation}
\sigma_r=\pi\bar{\lambda}^2\sum^\infty_{l=0}(2l+1)(1-\vert\eta_l\vert^2).
\label{sigr} \end{equation}
\begin{figure}
[t]
\begin{center}
\includegraphics[
natheight=1.5in, natwidth=1.5in, height=3.2in, width=3.2in
]%
{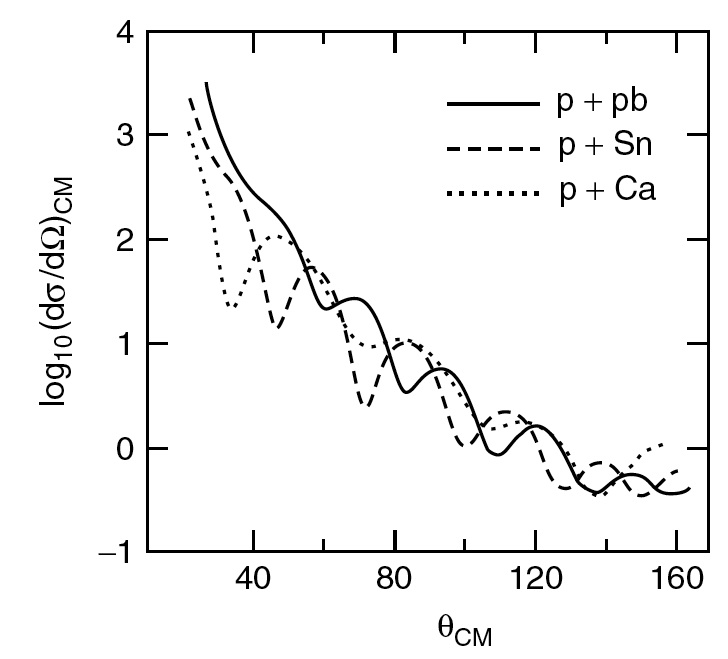}%
\caption{Elastic scattering angular distribution of protons of 30
MeV on {\rm Ca, Sn} and {\rm Pb.} The curves are
adjusted to the experimental data of ref. \cite{RT64}} %
\label{fig2}%
\end{center}
\end{figure}

From equations \ref{sige2} and \ref{sigr}: when $\vert\eta_l\vert=1
$ the reaction cross section is zero and we have pure scattering.
The contrary, however, cannot happen, as the vanishing of $\sigma_e
$ also implies in the vanishing of $\sigma_r $. In general there is
a region of allowed values of $\eta_l $ for which the two cross
sections can  coexist.

The maximum of $\sigma_r $ happens for $\eta_l=0 $, what corresponds
to total absorption. Let us suppose that the absorption potential is
limited to the surface of a nucleus with radius $R\gg\bar{\lambda}
$, that is, that all the particles with impact parameter smaller
than the radius $R $ are absorbed. That is equivalent to say that
all particles are absorbed for  $l\le R/\bar{\lambda} $. In this
case
\begin{equation}
\sigma_r=\pi\bar{\lambda}^2\sum_{l=0}^{R/\bar{\lambda}}(2l+1)=\pi(R+\bar{\lambda})^2.
\label{sigr2}
\end{equation}

This is the value that would be intuitively adequate for the total
cross section, i.e., equal to the geometric cross section (the part
$\bar{\lambda} $ can be understood as an uncertainty in the position
of the incident particle). But, we saw above that the presence of
scattering is always obligatory. For $\eta_l=0 $, the scattering and
reaction cross sections are identical, yielding the total cross
section
\begin{equation}
\sigma=\sigma_r+\sigma_e=\pi (R+\bar{\lambda})^2+\pi
(R+\bar{\lambda})^2=2\pi(R+\bar{\lambda})^2\label{sigr3}
\end{equation}
that is twice the geometric cross section!

The presence of the scattering part, that turns the result
\ref{sigr3} apparently strange, can be interpreted as the effect of
diffraction of the plane waves at the nuclear surface. This effect
leads to ``shadow'' behind the nucleus decreasing its apparent
diameter so that, at a certain distance, the perturbation caused by
the presence of the nucleus disappears and the plane wave is
reconstructed. In this situation we can say that the part of the
beam which is diffracted has to be the same as the part that is
absorbed, justifying the equality of $\sigma_r $ and $\sigma_e$. The
diffraction phenomenon appears clearly in the elastic scattering or
inelastic angular distribution (differential cross section as
function of the scattering angle): figure \ref{fig2} exhibits
angular distributions for the elastic scattering of 30 MeV protons
on $^ {40}$Ca, $^ {120}$Sn and $^ {208}$Pb. The oscillations in the
cross sections are characteristic of a {\it Fraunhofer diffraction}
figure, similar to light scattering by an opaque disk. The angular
distance $\Delta\theta $ between the diffraction minima follows
closely the expression $\Delta\theta=\hbar/pR $, characteristic of
diffraction phenomena.

\subsection{Excitation functions}

When the projectile has a very low energy, $k \rightarrow 0$, and in
particular, $l=kR \ll 1$. As an example consider the scattering of
neutrons with $l=0 $ and ignore the spins of the neutron and of the
target. In this case the {\it Schr\"odinger equation} for the radial
motion of the neutron with respect to the target is
\begin{equation}
{d^2u_0\over dr^2}+k^2u_0=0 \ \ (r\ge R)\label{neut1}.
\end{equation}
This is valid for the radial wave function  $u_0 $ at distances $r $
larger than the {\it channel radius} $R=R_a+R_A $, with $R_a $ and
$R_A $ being the radii of the projectile and of the target,
respectively. The solution of eq. \ref{neut1} is \begin{equation}
u_0=\eta_0e^{ikr}-e^{-ikr} \ \ \ \ \ \ \ \ (r\ge R). \label{neut2}
\end{equation}
A radial wave function inside the nucleus should
connect to the external function \ref{neut2} with a continuous
function and its derivative at $r=R $. That is, the function
\begin{equation}
f_l\equiv R\left[{du_l/dr\over u_l}\right]_{r=R}\label{neut3}
\end{equation}
must have identical values if calculated with the  internal or the
external function and this condition creates a relationship between
$f_l $ and $\eta_l $. Hence, the knowledge of $f_l $ leads to the
knowledge of the cross sections. For neutrons with $l=0 $, the
application of  \ref{neut2} results in
\begin{equation}
f_0=ikR\,{\eta_0+e^{-2ikR}\over\eta_0-e^{-2ikR}}, \label{neut4}
\end{equation}
from which we extract \begin{equation} \eta_0={f_0+ikR\over
f_0-ikR}e^{-2ikR}. \label{neut5}
\end{equation}

If $f_0 $ is a real number, then $\vert\eta_0\vert^2=1 $. The
reaction cross section, eq. \ref{sigr}, will be  zero and we have
pure scattering.

Using eq. \ref{neut5}, the scattering cross section, eq.
\ref{sige2}, can be written as
\begin{equation}\sigma_{e,0}=\pi\bar{\lambda}^2\vert
A_{res}+A_{pot}\vert^2,\label{sige3} \end{equation}  with $
A_{res}={-2ikR\over f_0-ikR}\label{ares} $ and $
A_{pot}=\hbox{exp}(2ikR)-1.\label{apot} $ The separation of the
cross section in two parts has physical justification: $A_{pot} $
does not contain $f_0 $ and therefore does not depend on conditions
inside the nucleus. It represents the situation where the projectile
does not  penetrate the nucleus as in the idealized situation where
the nucleus is considered an impenetrable hard sphere. The wave
function is zero inside the nucleus and $u_0 $ vanishes at $r=R $,
implying $f_0\rightarrow\infty $ and $A_{res}\rightarrow 0 $. Hence,
$A_{pot} $ is the only responsible for the scattering.

Inserting eq. \ref{neut5} in \ref{sigr} and using
\begin{equation}f_0=f_R+if_I\label{f0} \end{equation} we have
\begin{equation}\sigma_{r,0}=\pi\bar{\lambda}^2{-4kRf_I\over f_R^2+(f_I-kR)^2} \, \label{sigr0}\end{equation}
an equation that is useful when we study the presence of resonances
in the {\it excitation function} (cross section as a function of the
energy).
\bigskip
\section{Statistical reactions}

\subsection{Compound nucleus} When a low energy neutron ($<50 $ MeV)
enters the range of nuclear forces it can be scattered or  begin a
series of collisions with the nucleons. The products of these
collisions, including the incident particle, will continue in their
course, leading to new collisions and new changes of energy. During
this process one or more particles can be emitted and they form with
the residual nucleus the products of a reaction that is known as
{\it pre-equilibrium}. At low energies, the largest probability is
the continuation of the process so that the initial energy is
distributed among all nucleons, with no emitted particle. The final
nucleus with $A+1 $ nucleons has an excitation energy equal to the
kinetic energy of the incident neutron plus the binding energy the
neutron has in the new, highly unstable, nucleus \cite{Bo36}. It
can, among other processes, emit a neutron with the same or smaller
energy to the one absorbed. The de-excitation process is not
necessarily immediate and the excited nucleus can live a relatively
long time. We say that there is, in this situation, the formation of
a {\it compound nucleus} as intermediary stage of the reaction. In
the final stage the compound nucleus can evaporate one or more
particles, fission, etc. In our notation, for the most common
situation in which two final products are formed (the evaporated
particle plus the residual nucleus or two fission fragments, etc.)
we write:
$$\hbox{a}+\hbox{A}\rightarrow \hbox{C}^*\rightarrow
\hbox{B}+\hbox{b},\label{cnast} $$ the asterisk indicating that the
compound nucleus C is in an excited state.

\begin{figure}
[t]
\begin{center}
\includegraphics[
natheight=1.5in, natwidth=1.5in, height=3.2in, width=3.2in
]%
{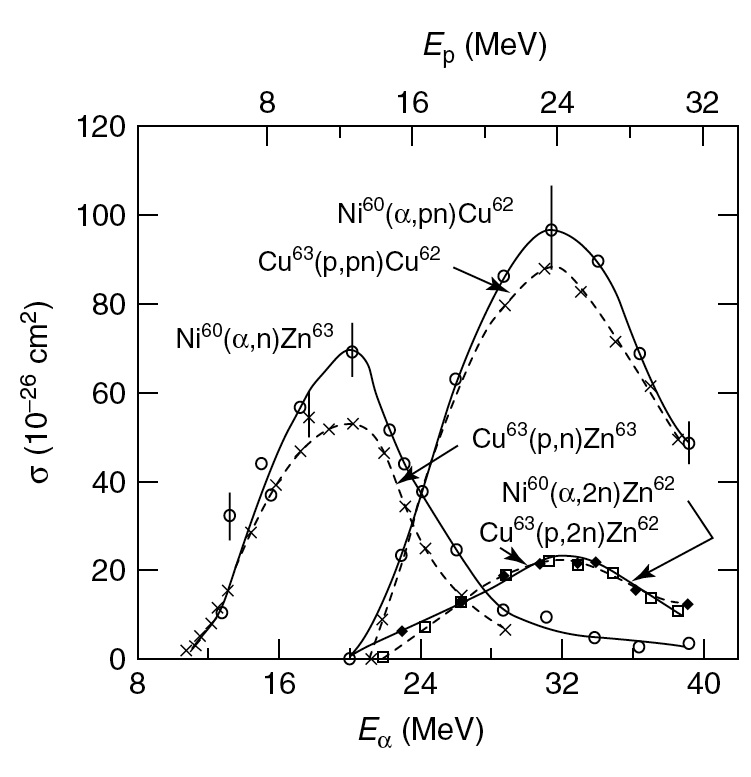}%
\caption{ Cross sections for the reactions shown in \ref{cneq}. The
scales of the upper axis (energy of the protons) and lower axis
(energy of the $\alpha $-particle) were adjusted  to correspond to
the same excitation energy of the compound nucleus \cite{Gh50}.}
\label{fig3}%
\end{center}
\end{figure}

The compound nucleus lives long enough  to ``forget'' how it was
formed and the de-excitation to the final products b and B only
depends on the energy, angular momentum and parity of the quantum
state of the compound nucleus. An interesting experimental
verification was accomplished by S. N. Ghoshal in 1950 \cite{Gh50}.
He studied two reactions that take to the same compound nucleus, $^
{64}$Zn$^ * $, and measured the cross sections of three different
forms of decay, as shown below:

\begin{eqnarray}
\hbox{p }+ \ ^ {63 }_ {} \hbox{Cu} \searrow \ \ \ \ \ \ \ \ \ \
\nearrow & \ \ ^ {63}\hbox{Zn }+\hbox {n} \nonumber \\ \ \ \ \ \ \ ^
{64}\hbox{Zn}^*\rightarrow & \ ^ {62}\hbox{Cu }+\hbox {n + p} \nonumber \\
a \ + \ ^ {60}\hbox{Ni} \nearrow \ \ \ \ \ \ \ \ \ \ \searrow & \ \
^ {62}\hbox{Zn} \ + \ \hbox{2n} \label{cneq}
\end{eqnarray}  If the idea of the
compound nucleus is valid and if one chooses the energy of the
proton and of the incident $\alpha$-particle to produce the same
excitation energy, then the cross section for each one of the three
exit channels should be independent of the way the compound nucleus
is formed. That is, the properties of the compound nucleus do not
have any relationship with the nuclei that formed it. This is
confirmed in figure \ref{fig3}, where one sees clearly that the
cross sections depend practically only on the exit channels.

The angular distribution of fragments, or evaporated particles, of a
compound nucleus  should be isotropic in the center of mass, and
this is verified experimentally. However, the total angular momentum
is conserved and cannot be ``forgotten''. Reactions with large
transfer of angular momentum, as  when heavy ions are used as
projectiles, can show a non-isotropic angular distribution in the
center of mass system.

The occurrence of a nuclear reaction in two stages allows the cross
section for a reaction A(a,b)B to be written as the product,
$\sigma(\hbox{a,b}) = \sigma_c(\hbox{a,A})P(\hbox{b}), $ where
$\sigma_c(a,A) $ is the cross section of formation of the compound
nucleus starting from the projectile a and  the target A and $P$(b)
is the probability that the compound nucleus emits
 a particle b leaving a residual nucleus B.
If the quantum numbers of entrance and exit channels are well
specified,  i.e., if the reaction begins at an entrance channel
$\alpha $ and ends at an exit channel $\beta $, one can write
\begin{equation}\sigma(\alpha,\beta)=\sigma_c(\alpha)P(\beta).
\label{alpha} \end{equation}

We can associate the probability $P(\beta) $ to the width
$\Gamma_\beta $ of the channel $\beta$ and write:
 \begin{equation}P(\beta)={\Gamma(\beta)\over\Gamma},\label{pbeta} \end{equation}
where $\Gamma $ is the total width, that is, $\tau=\hbar/\Gamma $ is
the half-life of disintegration of the compound nucleus. Eq.
\ref{pbeta} just expresses the fact that the decay probability
through channel $\beta$ is the decay rate through that channel
divided by the total decay rate. In the competition between the
several channels $\beta $, the nucleons have clear preference over
the $\gamma$-radiation whenever there is available energy for their
emission and among the nucleons the neutrons have preference as they
do not have the Coulomb barrier as an obstacle. Thus, in a reaction
where there is no restriction for neutron emission we can say that
\begin{equation}\Gamma\cong\Gamma_n,\label{gamma} \end{equation}
where $\Gamma_n $ includes the width for the emission of one or more
neutrons.

The study of the function $P(\beta) $ is done in an evaporation
model that leads to results in many aspects similar to the
evaporation of molecules of a liquid, with the energy of the emitted
neutrons having the form of a Maxwell-Boltzmann distribution
\begin{equation}
I(E)\propto E\exp\Bigl(-{ E\over\theta}\Bigr)dE,\label{iofe}
\end{equation} with $I $ measuring the amount of neutrons emitted
with energy between $E$ and $E+dE $. The quantity $\theta $, with
dimension of energy, has the role of a {\it nuclear temperature}. It
is related to the density of levels $\omega $ of the daughter
nucleus B by \begin{equation} {1\over\theta}={dS\over
dE},\label{dsde}
\end{equation} with
\begin{equation} S=\ln  \omega(E)\label{slne} \end{equation}
where $dS/dE $ is calculated for the daughter nucleus B at the
maximum excitation energy that it can have after the emission of a
neutron. That is, in the limit of emission of a neutron with zero
kinetic energy.
\begin{figure}
[t]
\begin{center}
\includegraphics[
natheight=1.2in, natwidth=1.2in, height=2.in, width=2.4in
]%
{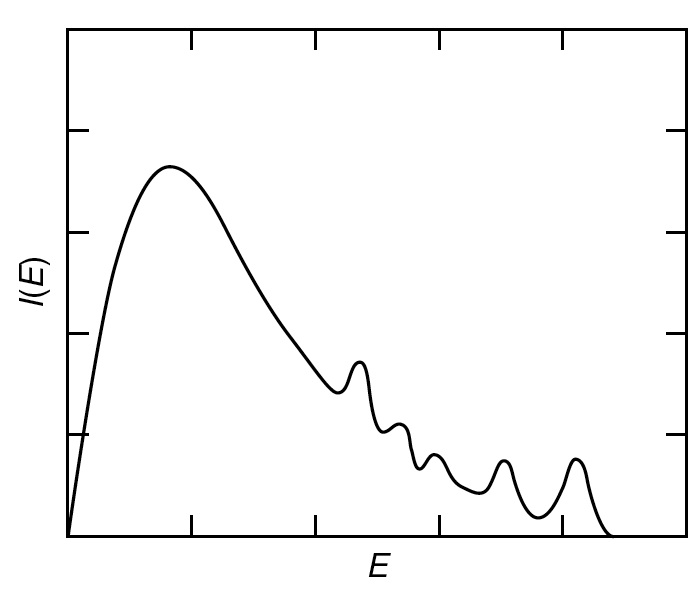}%
\caption{Energy spectrum of neutrons evaporated by a compound
nucleus.}
\label{fig4}%
\end{center}
\end{figure}

The level density $\omega(E) $ is a measure of the number of
available energy states for the decay of the compound nucleus in the
interval $dE $ around the energy $E $. In this sense, the
relationship \ref{slne} is, neglecting the absence of the Boltzmann
constant, identical to the thermodynamic relationship between the
entropy $S $ and the number of states available for the
transformation of a system. Eq. \ref{dsde} is the well-known
relation between the entropy and the temperature.

\begin{figure}
[ptb]
\begin{center}
\includegraphics[
natheight=14.5in, natwidth=16.1in, height=2.7in, width=3.0in ]
{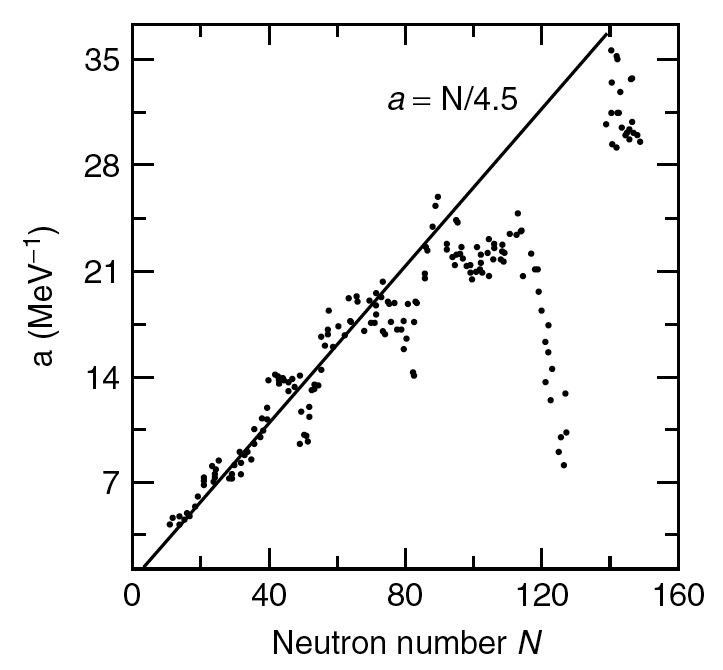}\caption{Values of the level density
parameter a as a function of the neutron number. Far from magic
regions, $a\approx N/4.5$ MeV$^{-1}$ which approximately corresponds
to $a=A/7.5$ MeV$^{-1}$ \cite{Fa68}. }
\label{nlevel}%
\end{center}
\end{figure}

A simple expression for the energy dependence of the state density
is provided by the equidistant spacing model which assumes that the
one-particle states are equally spaced with spacing $d$, and that
the total energy of the nucleus is simply obtained by adding the
energies of the constituent nucleons. The solution of this problem
can be obtained from statistical mechanics \cite{Fe924}:
\[
\rho\left(  E\right)  \sim\exp\left(  2\sqrt{aE}\right)  \ .
\]
Extensive analyzes of experimental data show that for nuclei far
from the region of the magic nuclei $a$\ varies linearly with $A$\
(or with $N$\ and $Z$), as shown in fig. \ref{nlevel}:
\begin{equation}
a\cong\frac{A}{k}\;\text{MeV}^{-1}.\label{16.27}%
\end{equation}
It is found that $k\cong7.5-8$.

\subsection{Energy spectrum of neutrons} The energy distribution of neutrons emitted by a compound
nucleus has the aspect of the curve shown in figure \ref{fig4}. Only
the low energy part obeys \ref{iofe} and the reason is simple: the
emission of a low energy neutron leaves the residual nucleus with a
large excitation energy, and the level density is very high. The
large density of final states turns the problem tractable with the
statistical model that leads to \ref{iofe}. In the opposite
situation are the low energy states of the residual nucleus. These
isolated states appear as peaks in the tail of the distribution.
When the emission is of a proton or of another charged particle, the
form of figure \ref{iofe} is distorted, the part of low energy of
the spectrum being suppressed partially by the Coulomb barrier.

\begin{figure}
[t]
\begin{center}
\includegraphics[
natheight=1.2in, natwidth=1.2in, height=2.in, width=3.4in
]%
{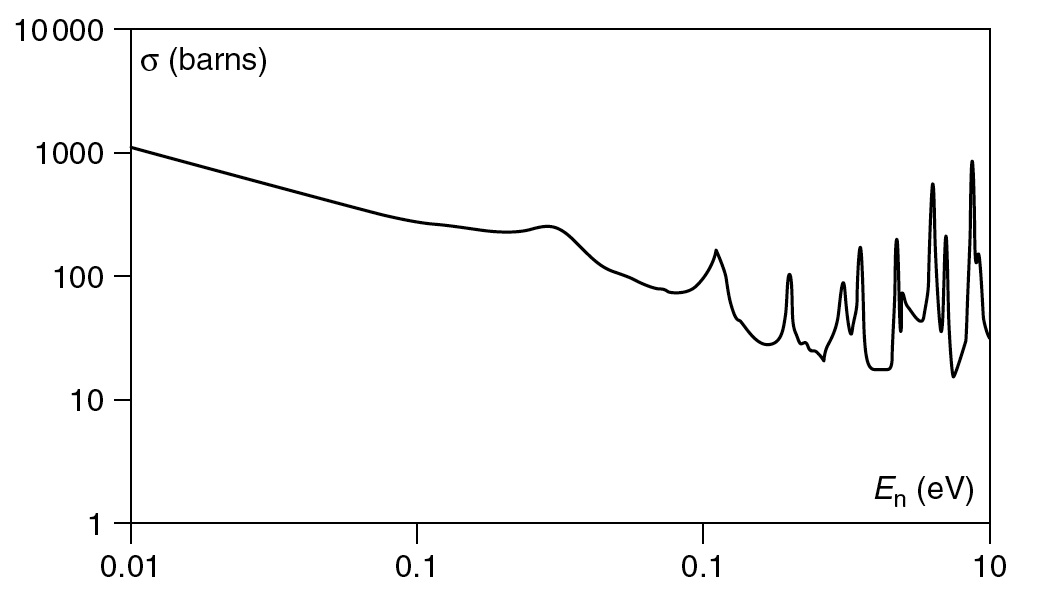}%
\caption{Total cross section for low energy neutrons hitting
$^{235}$U.}
\label{fig5}%
\end{center}
\end{figure}

Let us now assume that there is only elastic scattering or formation
of compound nucleus.  We further assume that the elastic scattering
is purely potential, without resonant elastic scattering. That is,
there is no re-emission of neutrons with energy equal to the
incident projectile. This is equivalent to say that the probability
that the exit channel is the same as the entrance channel  is very
low. According to these assumptions, the cross section
$\sigma_c(\alpha) $ of eq. \ref{alpha}  is the reaction cross
section of eq. \ref{sigr}. We can still write the wave function
inside the nucleus as just an incoming wave \begin{equation}u_0\cong
\hbox{exp}(-iKr) \ \ \ \ \ \ \ \ (r < R),\label{uofo}\end{equation}
where $K=\sqrt{2m(E-V_0)} / \hbar $ is the wavenumber inside the
nucleus, and it is assumed that the neutron with total energy $E $
is subject to a negative potential $V_0 $. Eq. \ref{uofo} is clearly
a crude simplification when the incident neutron interacts in a
complicated way with the other nucleons in the nucleus. It allows,
however, to explain the average behavior of the cross sections for
low energies. Starting with eqs. \ref{neut3} and \ref{uofo} one
determines the value of $f_0 $:
\begin{equation}f_0=-iKR\label{fokr}\end{equation} and from eq.
\ref{sigr0}, we get
\begin{equation}\sigma_c=\pi\bar{\lambda}^2{4kK\over(k+K)^2} \label{sigc}\end{equation}
for the cross section of compound nucleus formation for neutrons
with $l=0 $. At low energies, $E << \vert V_0\vert $, thus $k << K
$. Under these conditions, $\sigma_c=4\pi/kK $, since
$\bar{\lambda}=k^{-1}
>> R$. Thus, $\sigma_c $ varies with $1/k $. That is,
\begin{equation}\sigma_c\propto{1\over v},\label{oneov} \end{equation}
where $v $ is the velocity of the incident neutron. This is the
well-known $1/v $   {\it law} that governs the behavior of the
capture cross section of low energy neutrons. Figure \ref{fig5}
exhibits the excitation function (cross section as function of the
energy) for the reaction n+$^{235}$U. The cross section decays with
$1/v $ up to 0.3 eV, where a series of resonances appear.

\subsection{Resonances} To understand why there are resonances, we
shall use again the simple model of a single particle subject to a
square-well potential. We know that inside the well the
Schr\"odinger equation only admits solution for a discrete group of
values of energy, $E_1 $, $E_2 $,... $E_n $. A particle is confined
to the interior of the well by reflections that it has at the
surface of the well. In these reflections the wave that represents
the particle should be in phase before and after the reflection and
this only happens for a finite group of energies. Outside the well
the Schr\"odinger equation does not impose restrictions and the
energy can have any value. But we know, from the study of the
passage of a beam of particles through a potential step, that the
discontinuity of the potential at the step provokes reflection even
when the total energy of the particles is larger than the step, a
situation where classically there would not be any difficulty for
the passage of the particles. This reflection is partial and it
becomes larger the closer the energy is closer to the height of the
potential step. We can say that a particle with energy slightly
positive is almost as confined as a particle inside the well. From
this fact results the existence of almost bound states of positive
energy known as {\it quasi-stationary states} or {\it resonances}.
These resonances appear as peaks in the excitation function, a peak
at a given energy meaning that the energy coincides with a given
resonance of the nucleus.

The existence of resonances can also be inferred from the properties
of the wave function. We consider only elastic scattering, with the
other channels closed. The external and internal wavefunctions are
both sine functions, the first with wavenumber $k=\sqrt{2mE}/\hbar $
and the second with $K=\sqrt{2m(E-V_0)} / \hbar $. If $E $ is small
and $V_0 $ is about $-35$~MeV, we have $K
>>k $. The internal and external parts should join at $r=R $ with
continuous function and derivatives. As the internal frequency is
much larger than the external one, the internal amplitude is quite
reduced. Only at the proximity of the situation in which the
derivative is zero there is a perfect matching between both and the
internal amplitude is identical to  the external one. The energy for
which this happens is exactly the energy of resonance.

Resonances appear in the excitation function at relatively low
energies, where the number of open channels is not very large and it
is possible that to return to the entrance channel. To arrive at an
expression of the cross section that describes a resonance, we
rewrite \ref{uofo} as,
\begin{equation}u_l\sim \hbox{exp}(-iKr)+b \ \hbox{exp}(iKr)\; , \ \ \ \
\ \ \ \ (r <R),\label{uofo1}\end{equation} {this time containing a
second part which takes into consideration the part of the wave that
returns. This second part allows the existence of resonant
scattering, where the incident particle is re-emitted with the same
energy that it entered, after forming the compound nucleus. The
complex amplitude $b $ is always  smaller than one, because there
are no creation of particles in the region $<R$ in eq. \ref{sigr0}.

The second parenthesis in the denominator of eq. \ref{sigr0} is
never zero, since the numerator forces $f_I $ to be always negative.
If for  a certain energy $f_R $ vanishes, $\sigma_{r,0} $ passes by
a maximum in that energy. We can tentatively identify these energies
as being the energy of the resonances. Let us take the extreme case
of a single resonance at the energy $E_R $, that is, $f_R=0 $ for
$E=E_R $. We can expand $f_R $ in a Taylor series in the
neighborhood of a resonance, $f_R(E)=(E-E_R)({df_R/
dE})_{E=E_R}+...$.  Keeping just the first term of the expansion and
using \ref{sige3} and \ref{sigr0}, we get
\begin{equation}\sigma_{e,0}=\pi\bar{\lambda}^2\Bigl\vert
\hbox{exp}(2ikR)-1+{i\Gamma_\alpha\over(E-E_R)+i{\Gamma\over2}}
\Bigr\vert^2,\label{sige02}\end{equation}
\begin{equation}\sigma_{r,0}=\pi\bar{\lambda}^2
{\Gamma_\alpha(\Gamma-\Gamma_\alpha)\over(E-E_R)^2+\left({\Gamma\over2}\right)^2},\label{sigr02}\end{equation}
where we define: \begin{equation}\Gamma_\alpha=-{2kR\over
(df_R/dE)_{E=E_R}} \ \quad\hbox{and}\quad\Gamma={2kR-2f_I\over
(df_R/dE)_{E=E_R}}. \label{gam03}\end{equation}

The energy $\Gamma $ that appears in \ref{sige02}, is the total
width of the resonance, $\Gamma=\Gamma_\alpha+\Gamma_\beta+... $,
i.e., the sum of the widths for all the possible processes of decay
of the nucleus, starting from the resonant state. $\Gamma_\alpha $
is the entrance channel width, and $\Gamma-\Gamma_\alpha $ is the
sum of the widths of all the exit channels except $\alpha $. If we
restrict the exit channels to a single channel $\beta $, or we
denote  $\beta $ as the group of exit channels except $\alpha $, eq.
\ref{sigr02} is rewritten as \begin{equation}\;\;\;
\sigma_{\alpha,\beta} =\pi\bar{\lambda}^2{\Gamma_\alpha\Gamma_\beta
\over(E-E_R)^2+\left({\Gamma\over2}\right)^2},\label{BW1}\end{equation}
which is the usual way of presenting the {\it Breit-Wigner formula},
which describes the form of the cross section close to a resonance.
Let us recall that eq. \ref{BW1}  refers to an incident particle of
$l=0 $, without charge and without spin. If the spins of the
incident and target particles are $s_a $ and $s_A$, respectively,
and the incident beam is described by a single partial wave $l\ne0
$, one can show that the cross section of eq. \ref{BW1} should be
multiplied by the statistical factor $g=(2I+1)/(2s_a+1)(2s_A+1)$
where $I$ is the quantum number of the total angular momentum
$\bf{I}={\bf s}_a+{\bf s}_A + {\bf l}$ of the compound nucleus. $g$
reduces, naturally, to the unit in the case of zero intrinsic and
orbital angular momenta.

\begin{figure}
[t]
\begin{center}
\includegraphics[
natheight=1.2in, natwidth=1.2in, height=3.2in, width=2.8in
]%
{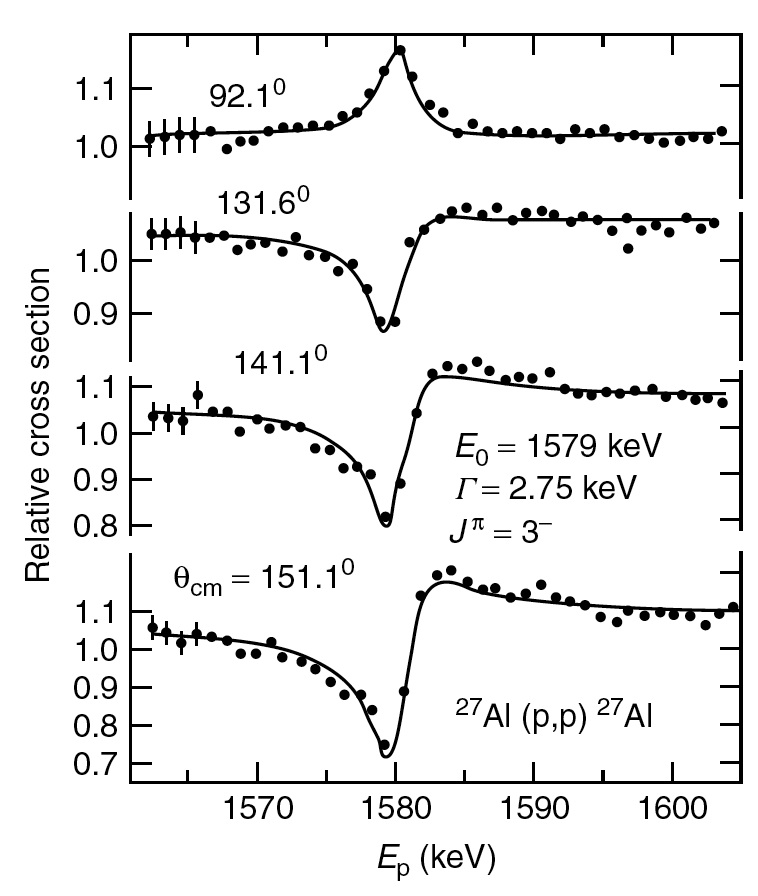}%
\caption{Differential cross section at four angles for the elastic
scattering of protons off $^ {27}${\rm Al}, in units of the
Rutherford cross section, in the neighborhood of the 1579 keV
resonance. For larger angles we have typical interference  between
the resonant scattering and the potential scattering \cite{Tv72}.}
\label{resonance}%
\end{center}
\end{figure}

If the exit channel is the same as the entrance channel $\alpha $,
the cross section should be obtained from \ref{sige02} and its
dependence in energy is more complicated because in addition to the
resonant scattering there is the potential scattering, and the cross
section  \ref{sige02} will contain  an interference term between
both. The presence of interference results in a peculiar aspect of
the scattering cross section, which differs from the simple form
\ref{BW1} for the reaction cross section. This is seen in figure
\ref{resonance} that shows the forms that a resonance can take in
the scattering cross section.

The region of energy where resonances show up can extend to 10 MeV
in light nuclei but it ends well before this in heavy nuclei.
Starting from this limit the increase in level density with energy
implies that the average distance between levels is smaller than the
width of the levels and individual resonances cannot be resolved
experimentally. They form a continuum and this region is known as
{\it continuum region}. The cross-section in this region fluctuates.

\begin{figure}
[t]
\begin{center}
\includegraphics[
natheight=1.2in, natwidth=1.2in, height=2.8in, width=2.6in
]%
{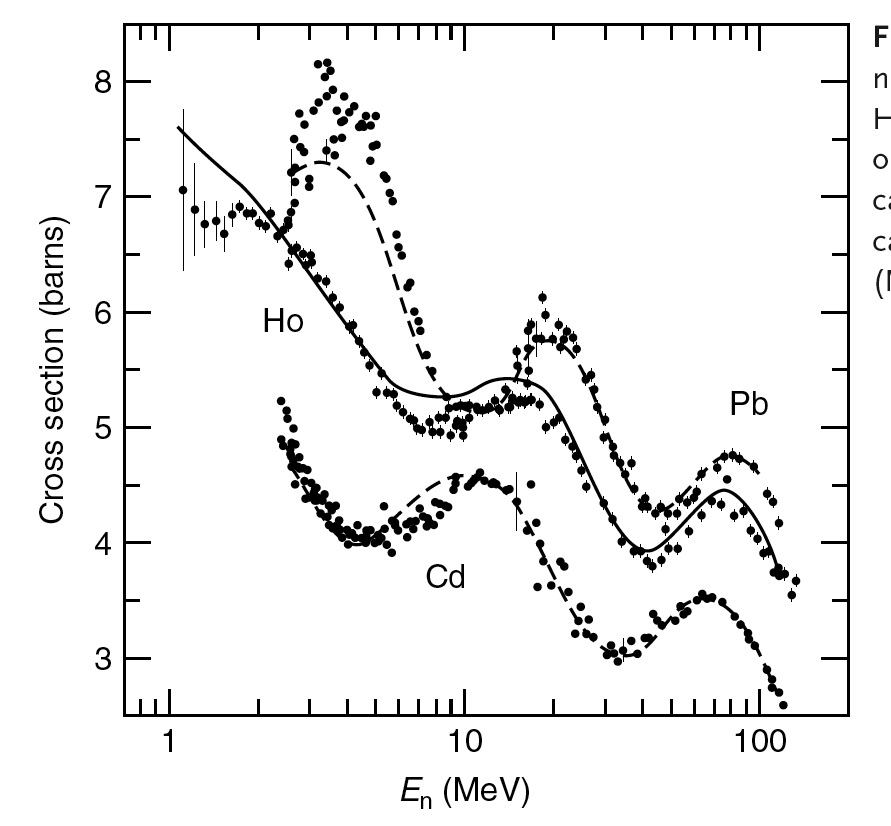}%
\caption{Total cross section for neutrons incident on cadmium,
holmium and lead, showing an oscillatory behavior. The curves for
cadmium and lead are results of calculations with the optical model
\cite{Ma68}.}
\label{nxc}%
\end{center}
\end{figure}

\subsection{Statistical theory of nuclear reactions}
The fluctuating behavior of low-energy nuclear reactions is due to
the interference of the reaction amplitudes corresponding to the
excitation of each of the overlapping states which vanish in the
energy average of the cross-section since these amplitudes are
complex functions with random modulus and phase. Calling
$\sigma_N(c)$ the cross section for the formation of a compound
nucleus in the entrance channel $c$, and using the
\textit{reciprocity theorem} which relates the cross-section
$\sigma_{cc^{\prime}}$ to the cross-section for the time-reversed
process $c^{\prime}\rightarrow c$, one gets
\begin{eqnarray}
&&\sigma_{cc^{\prime}}\left(  E_{c^{\prime}}\right)
dE_{c^{\prime}}=\sigma _{CN}\left(  c\right) \nonumber \\
&\times& \frac{\left( 2I_{c^{\prime}}+1\right)  \mu_{c^{\prime
}}E_{c^{\prime}}\;\sigma_{CN}\left(  c^{\prime}\right)  \omega(U_{c^{\prime}%
})dU_{c^{\prime}}}{\sum_{c}\int_{0}^{E_{c}^{\max}}\left(
2I_{c}+1\right)
\mu_{c}E_{c}\;\sigma_{CN}\left(  c\right)  \omega(U_{c})dU_{c}}%
\ ,\label{21.54b}%
\end{eqnarray}
where $I_{c}$ is the angular momentum and $\mu_{c}$ is the reduced
mass in channel $c$. Ejectiles with energy in the range $E_{c^{\prime}}$ to $E_{c^{\prime}%
}+dE_{c^{\prime}}$ leave the residual nucleus with energy in the
range $U_{c^{\prime}}$ to $U_{c^{\prime}}+dU_{c^{\prime}}$ where
$U_{c^{\prime}}=E_{CN}-B_{c^{\prime}}-E_{c^{\prime}}$ and $E_{CN}$
and $B_{c^{\prime}}$ are respectively the compound nucleus energy
and the binding energy of the ejectile in the compound nucleus. Eq.
\ref{21.54b} is the \textit{Weisskopf-Ewing formula} for the
angle-integrated cross-sections \cite{WW40}. To a good
approximation, the level density $\omega(U)\propto\exp(U/T)$, so the
ejectile spectrum given by the Weisskopf-Ewing theory is Maxwellian.
It rises rapidly above the threshold energy, attains a maximum and
then falls exponentially.

The Weisskopf-Ewing theory simple to use, but it has the
disadvantage that it does not explicitly consider the conservation
of angular momentum and does not give the angular distribution of
the emitted particles. This is provided by the
\textit{Hauser-Feshbach theory} \cite{HF52}. This theory takes into
account the formation of the compound nucleus in states of different
$J$ and parity $\pi$. Let us consider the case of a reaction leading
from the initial channel $c$ to a final channel $c^{\prime}$. If
there is no {\it pre-equilibrium emission}, one may identify the
compound nucleus formation cross-section $
\sigma_{CN}=\sum_{J,\pi}\sigma_{CN}^{J\pi}\label{21.68}%
$
with the optical model reaction cross-section
$
\sigma_{R}=\frac{\pi}{k^{2}}\sum_{l}\left(  2l+1\right)  T_{l}\ ,\label{21.69}%
$
which, if the transmission coefficients $T_{l}=1-\left\vert
\left\langle S_{l}\right\rangle \right\vert ^{2}$ do not depend on
$J$. $\left\langle S_{l}\right\rangle $ is the average value of the
scattering amplitude over several overlapping resonances.

Using the above assumptions, the reciprocity theorem, and following
a similar derivation as with the Ewing-Weisskopf theory, the
cross-section for transition $c\rightarrow c'$ is given by the
\textit{Hauser-Feshbach} expression
\begin{equation}
\sigma_{cc^{\prime}}=\frac{\pi}{k^{2}}\sum_{J}\frac{\left(
2J+1\right)
}{\left(  2i_{c}+1\right)  \left(  2I_{c}+1\right)  }\frac{\sum_{s,l}%
T_{l}(c)\sum_{s^{\prime},l^{\prime}}T_{l^{\prime}}(c^{\prime})}{\sum_{c}%
\sum_{s,l}T_{l}(c)}\ .\label{21.73}%
\end{equation}

The compound nucleus states may be both of positive and negative
parity. Since parity is conserved, in evaluating (\ref{21.73}), one
must take into account that the parity of compound nucleus states
and the parity of the residual nucleus states may impose
restrictions to the values of the emitted particle angular momentum.
Thus, positive parity compound nucleus states decay to positive
parity states of the residual only by even angular momenta and to
negative parity residual nucleus states by odd angular momenta.

\subsection{The optical model}

Expression \ref{sigr3} shows that if a particle is always absorbed
when it reaches the nucleus, the total cross section  falls
monotonically with the energy and grows linearly with $A^{1\over3}
$. In figure \ref{nxc} we see curves of total cross section of
neutron scattering showing an oscillatory behavior for its energy
dependence as well as for the mass dependence of the target.  Their
presence is mainly due to interference phenomena between the part of
the incident beam that passes through the nucleus and the part that
passes around it \cite{Mv67}.

The basis of the {\it optical model} was established by Herman
Feshbach and collaborators in 1953 \cite{Fe53}. In this model the
interaction between the nuclei in a reaction is described by a
potential $U(r) $, with $r $ being the distance between the center
of mass of the two nuclei.  One replaces the complicated interaction
that a nucleon has with the rest of the nucleus with a potential
that acts on the nucleon. The potential $U(r) $ includes a complex
part that takes into account the absorption effects, i.e., the
inelastic scattering. The nuclear scattering is treated in similar
form as the scattering of  light by a glass sphere and the name of
the model derives of this analogy.

In its most commonly used form, the optical potential is written as
the sum:
\begin{equation}U(r)=U_R(r)+U_I(r)+U_D(r)+U_S(r)+U_C(r),\label{uopt1}\end{equation}
which contains parameters that can vary with energy and masses of
the nuclei and that should be chosen by an adjustment to the
experimental data. Obviously, the optical potential $U(r) $ will
only make sense if these variations are small for neighboring
 masses or energies.

The first part of \ref{uopt1}, $U_R(r)=-Vf(r,R,a),$ is real and
represents a nuclear well with depth $-V $, being multiplied by a
Woods-Saxon form factor
$f(r,R,a)=\left[1+\exp\{(r-R)/a\}\right]^{-1},$ where $R $ is the
radius of the nucleus and $a $ measures the diffuseness of the
potential, i.e., the width of the region where the function $f $ is
sensibly different from 0 or 1. $V $, $R $ and $a $ are treated as
adjustable parameters.

The absorption effect or, in another way, the disappearance of
particles from the elastic channel, is taken into account including
the two following imaginary parts $U_I(r)=-iWf(r,R_I,a_I)$ and
$U_D(r)=4ia_IW_Ddf(r,R_I,a_I)/dr$. An imaginary part produces
absorption. It is easy to see this for the scattering problem of the
square well: if an imaginary part is added to the well,
$U(r)=-V_0-iW_0\ \ \  (r <r_0)$ and $U(r)=0\ \ \ \  \ (r>r_0)$, it
appears in the value of $K=[2m(E+V_0+iW_0)] ^ {1/2}/\hbar $. This
will produce an exponentially decreasing internal wavefunction.
Thus, it corresponds to an absorption of particles from the incident
beam.

$U_I$ is responsible for the absorption in the whole volume of the
nucleus, but $U_D$, built from the derivative of the function $f $,
acts specifically in the region close to the nuclear surface. These
two parts have complementary goals: at low energies there are no
available unoccupied states for nucleons inside the nucleus and the
interactions are essentially at the surface. In this situation
$U_D(r) $ is important and $U_I(r) $ can be ignored. On the other
hand, at high energies  the incident particle has larger penetration
and in this case  the function $U_I(r) $  is important.
\begin{figure}
[t]
\begin{center}
\includegraphics[
natheight=1.2in, natwidth=1.2in, height=2.55in, width=2.3in
]%
{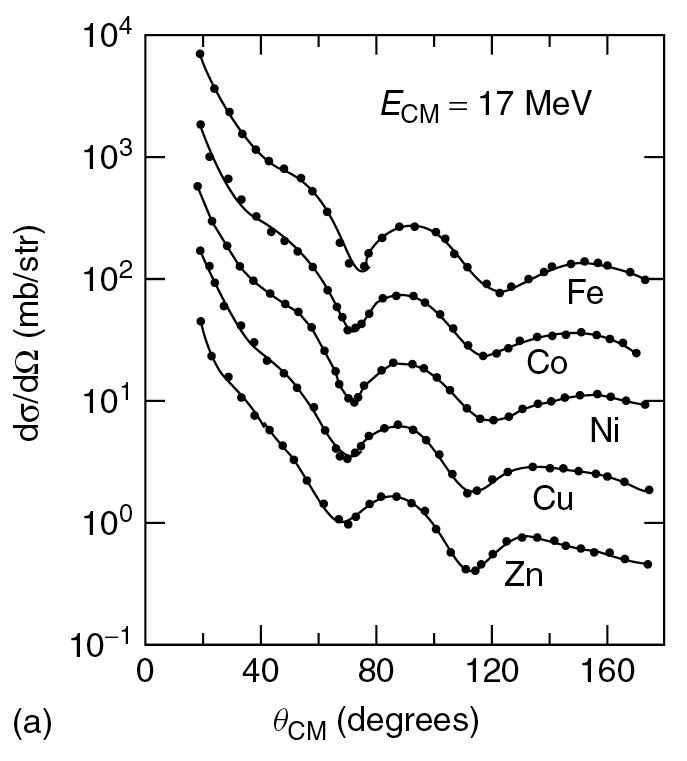}%

\smallskip

\includegraphics[
natheight=1.2in, natwidth=1.2in, height=2.3in, width=2.35in
]%
{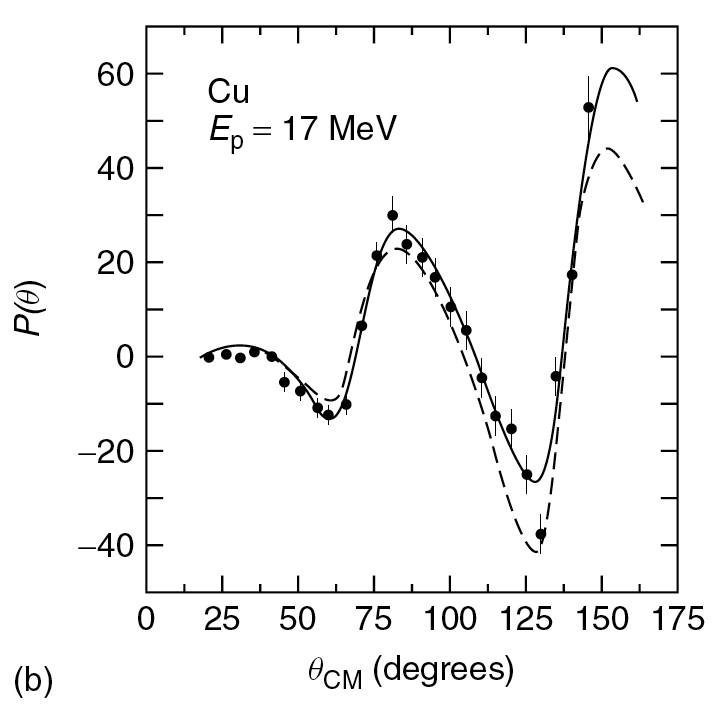}%
\caption{(Top) Angular distribution of the elastic scattering of 17
MeV protons on nuclei in the region $Z=26-30 $. (Bottom) The
polarization value \ref{pol} in the scattering of 9.4 MeV  protons
on copper. The curves are, in both cases, obtained with fits from
the optical model \cite{Pe63}.}
\label{optfig}%
\end{center}
\end{figure}
As with the shell model potential, used in nuclear structure, a
spin-orbit interaction term is added to the optical potential. This
term, which is the fourth part of \ref{uopt1}, is usually written in
the form \begin{equation}U_S(r)={\bf s\cdot\l}\;V_s\;{1\over
r}{d\over dr}f(r,R_S,a_S).\label{uopso}\end{equation} ${\bf s} $ is
the spin operator and ${\bf l} $ the angular orbital momentum
operator. As with $U_D(r) $, the part $U_S(r) $ is only important at
the surface of the nucleus since it contains the derivative of the
form factor $f$. The values of $V_S $, $R_S $ and $a_S $ must be
adjusted by experiment.

The presence of the term $U_S$ is necessary to describe the effect
of {\it polarization}. Through experiences of double scattering it
can be verified that proton or neutron beams suffer strong
polarization at certain angles. This means that the quantity
\begin{equation}P={N_c-N_b\over N_c+N_b},\label{pol}\end{equation}
where $N_c $ is the number of nucleons in the beam with spin up and
$N_b $ with spin down,  has a value significantly different from
zero at these angles. With the inclusion of $U_S$, the optical model
is able to reproduce in many cases the experimental values for the
polarization \ref{pol}.

Finally, a term corresponding to the Coulomb potential is added to
\ref{uopt1} whenever the scattering involves charged particles. It
has the form
\begin{eqnarray}U_C(r)&={Z_1Z_2e^2\over2R_c}\left(3-{r^2\over
R_c^2}\right)\qquad\qquad(r\le R_c)\\ & = {Z_1Z_2e^2\over
r}\qquad\qquad\qquad\qquad\quad(r > R_c),\label{ucoul}\end{eqnarray}
where is assumed that the nucleus is a  homogeneously charged sphere
of radius equal to the {\it Coulomb barrier radius} $R_c $, which
separates the regions of nuclear and Coulomb forces.

Figure \ref{optfig} exhibits the result of the application of eq.
\ref{uopt1} to the elastic scattering of 17~MeV protons on several
light nuclei. The angular distribution is very well reproduced by
the model, which also reproduces correctly the polarization
\ref{uopt1} for copper as a function of the scattering angle.

The optical model has a limited set of adjustable parameters and is
not capable to describe abrupt variations in the cross sections, as
it happens for isolated resonances. However, it can do a good
description of the cross sections in the presence of the
oscillations of large width in the continuous region, as it treats
these as wave phenomena.

\section{Direct reactions}
Direct reactions becomes more probable as one increases the energy
of the incident particle: the wavelength associated to the particle
decreases and localized areas of the nucleus can be ``probed'' by
the projectile. In this context,  peripheral reactions, where only a
few nucleons of the surface participate become important. Direct
reactions happen during a time of the order of 10$^{-22} $s.
Reactions with formation of  compound nuclei can be up to six orders
of magnitude slower. A reaction type at a given energy is not
necessarily exclusive; the same final products can be obtained, part
of the events in a direct way, other parts through the formation and
decay of a compound nucleus.

\begin{figure}
[t]
\begin{center}
\includegraphics[
natheight=1.2in, natwidth=1.2in, height=2.3in, width=3.5in
]%
{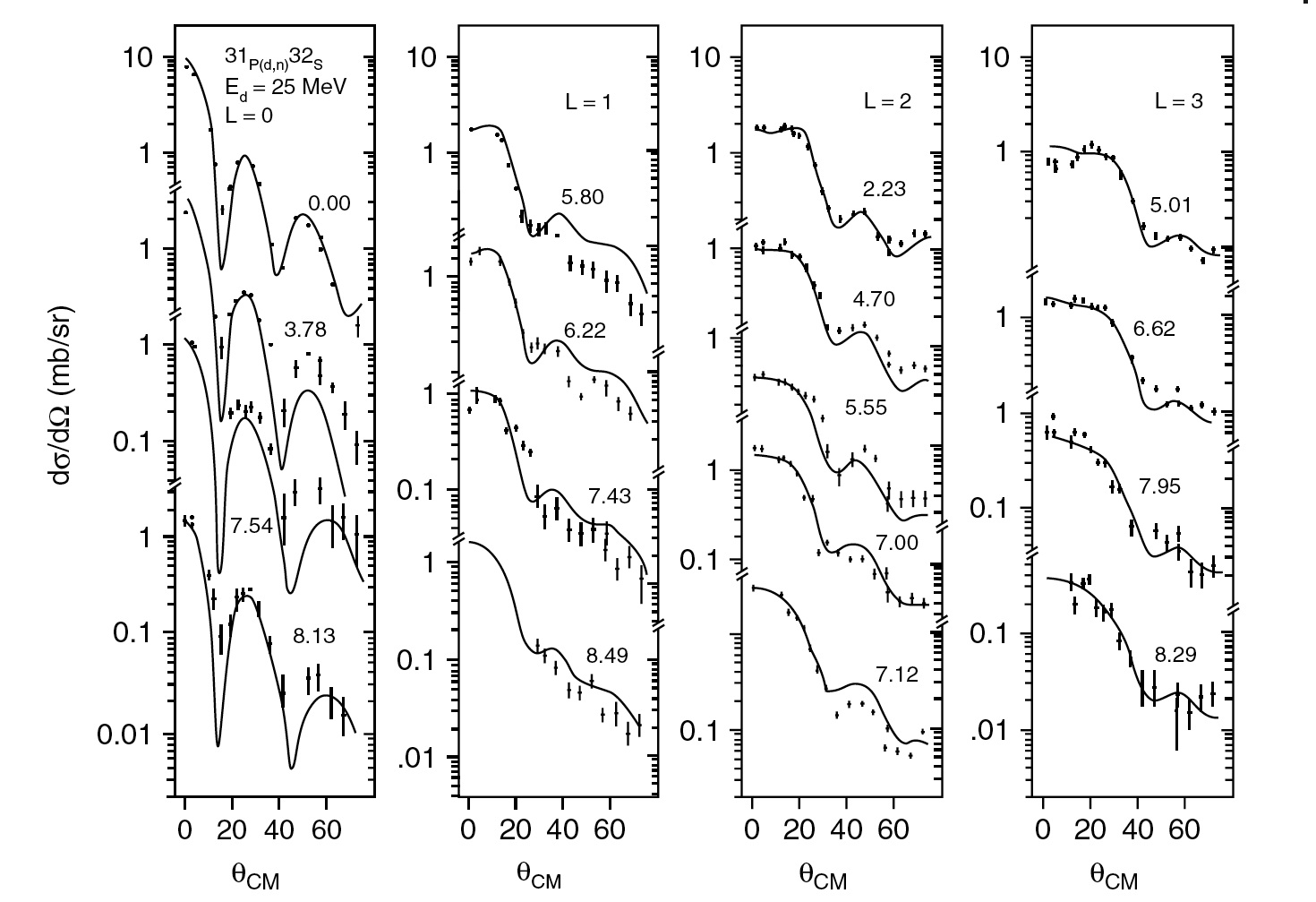}%
\caption{Angular distribution of the reaction
{\rm$^{31}$P(d,n)$^{32}$S}, with the transfer of a proton to several
states of $^ {32}${\rm S}. The curves are results of DWBA
calculations for the indicated $l $  values \cite{Mi87}.}
\label{dfig}%
\end{center}
\end{figure}

There are two characteristic types of direct reactions. In the first
the incident particle scatters inelastically and the transferred
energy is used to excite a collective mode of the nucleus.
Rotational and vibrational bands can be studied in this way. The
second type involves a modification in the nuclear composition.
Examples are transfer of nucleons, as {\it pick-up} and {\it
stripping} reactions. An important reaction of the latter kind is a
{\it knock-out} reaction where the incident particle knocks out a
particle of the target nucleus and continues in its path, resulting
in three reaction products. Reactions with nucleon exchange can also
be used to excite collective states. An example is a pick-up
reaction where a projectile captures a neutron from a deformed
target and the product nucleus is in an excited state belonging to a
rotational band.

Direct reactions exhibit a peculiar form of angular distribution,
which allows us to extract information on the reaction mechanism
with the employment of simple models. Typical examples are the
stripping reactions (d,n) and (d,p), where the angular distribution
of the remaining nucleon presents a forward prominent peak and
smaller peaks at larger angles, with the characteristic aspect of a
diffraction figure.

Figure \ref{dfig} shows experimental results for the reaction $^
{31}$P(d,n)$^{32}$S \cite{Mi87}: angular distributions of the
detected neutrons corresponding to each energy level of  $^ {32}$S
are exhibited for several angular momentum. We see that the behavior
of the cross sections is in agreement with qualitative predictions:
the curves exhibit a first peak at a value of $\theta $ that grows
with $l $. Other smaller peaks occur as $\theta $ increases. The
increase of $\theta_l $ with $l $ is an important characteristic
that can be used to identify the value of the transferred momentum
in a given angular distribution.

We consider an initial quantum state with particles of mass $m_a $
hitting a target A of mass $m_A $. The final quantum state are
particles of mass $m_b $ moving away from nucleus B of mass $m_B $.
The differential cross section for the process is given by
\begin{equation}{d\sigma\over
d\Omega}={m_am_bk_b\over(2\pi\hbar^2)^2k_a} \vert
V_{fi}\vert^2,\label{dir2}\end{equation} involving the matrix
element \begin{equation}V_{fi}=\int\Psi^*_b\Psi^*_B\Psi_\beta({\bf
r}_\beta) V\Psi_a\Psi_A\Psi_\alpha({\bf
r}_\alpha)d\tau.\label{dir3}\end{equation}
$\Psi_a,\Psi_b,\Psi_A,\Psi_B $ are the internal wavefunctions of the
nuclei a, b, A and B. $\Psi_\alpha,\Psi_\beta $ are the
wavefunctions of the relative motion in the entrance channel $\alpha
$ and in the exit channel $\beta $. The integration volume $d\tau$
spans the coordinates of all particles.  $V $ is the perturbation
potential that causes the ``transition'' from the entrance to the
exit channel.

The use of plane waves for $\Psi_\alpha $ and $\Psi_\beta $ in eq.
\ref{dir3} is known as {\it first Born approximation}. With it we
can arrive to an approximate expression for the behavior of the
differential cross section. As the nuclear forces are of short
range, one can restrict the integral \ref{dir3} to regions where
${\bf r}_\alpha\cong{\bf r}_\beta={\bf r} $. This leads to
\begin{equation}V_{fi}\cong\int d{\bf r}\exp(i{\bf k}\cdot{\bf r})
\biggl\{\int\Psi^*_b\Psi^*_B V \Psi_a\Psi_A\,d\tau'\biggr \},
\label{dir4}\end{equation} where {\bf k} = {\bf k}$_\alpha - ${\bf
k}$_\beta $. The global variables in $d\tau $ were separated into
variables $d{\bf r} $ and $d\tau ' $.

Expanding the plane wave in a Legendre polynomial series, we obtain
\begin{equation}V_{fi}\cong\sum_{l=0}^\infty i^l(2l+1)\int
j_l(kr)P_l(\cos\theta)F({\bf r}) d{\bf r},
\label{dir5}\end{equation} where $F({\bf r}) = \int\Psi^*_b\Psi^*_B
V \Psi_a\Psi_A d\tau ' $ contains all the internal properties and is
known as as the {\it form factor} of the reaction. The action of $V
$ is restricted to the surface of the nucleus: outside the nucleus
the action of $V $ is limited by the short range of the nuclear
forces and inside  the nucleus there is a strong deviation to the
absorption channel. Expression \ref{dir5} becomes
$V_{fi}\cong\sum_{l=0}^\infty c_l j_l(kR)$, where the coefficients
$c_l $ contain information on the form factor $F({\bf r}) $. The
index $l $ can be identified as the angular momentum transferred
and, for a reaction that involves a single value of $l $, we can
write for the differential cross section:
\begin{equation}{d\sigma\over d\Omega}\propto\vert
j_l(kR)\vert^2,\label{dir6}\end{equation} where the dependence in
$\theta $ is contained in
$$k^2=k_\alpha^2+k_\beta^2-2k_\alpha
k_\beta\cos\theta.\eqno(10.82)$$ We have an oscillatory behavior for
the angular distribution, the maxima separated by $\pi $ from each
other in the axis $kR $.

The Born approximation with plane waves predicts for certain cases
the correct place of the first peaks in the angular distribution but
without reproducing correctly the intensities. A considerable
progress can be done in the perturbative calculations if, instead of
plane waves in \ref{dir3}, we use {\it distorted} waves that
contain, besides the plane wave, the part dispersed elastically by
the optical potential of the target. The Born approximation with
distorted waves, or DWBA ({\it distorted wave Born approximation}),
became a largely employed tool in the analysis of experimental
results of direct reactions. With it one can try to extract with a
certain reliability the value of the angular momentum $l $
transferred to the nucleus in a stripping or pick-up reaction.  An
example of this is the  already mentioned stripping reaction, $^
{31}$P(d,n)$^{32}$S,  for deuterons of 25 MeV. For the  energy
levels  shown in figure \ref{dfig} the assignments of the value of
$l $ for the level is, in most cases, univocal.

The angular momentum $l $ transferred in a direct reaction generally
modifies the value of the total angular momentum of the nucleus. If
$J_i $ is the spin of the target nucleus, the spin $J_f $ of the
product nucleus  is limited to the values
\begin{equation}\Bigl\vert\bigl\vert
J_i-l\bigr\vert-{1\over2}\Bigr\vert\le J_f\le
J_i+l+{1\over2},\label{dir7}\end{equation} and the initial and final
parities obey the relationship $\pi_i\pi_f=(-1)^l.$ Eq. \ref{dir7}
allows, with the knowledge of the target nucleus and of the
transferred angular momentum, the determination of the parity of the
product state formed and is a tool for the determination of its
spin.

The knowledge of the  transferred angular momentum  value in a
direct reaction opens the possibility to test the predictions of the
shell model for the structure of nuclei. In a direct reaction one
assumes that the nucleon is located in an orbit of the nucleus with
the same  angular momentum as the transferred momentum in the
reaction. In almost all cases studied with direct reactions the
value of the assigned $l$ corresponds exactly to the predicted by
the shell model. We know, however, that the real situation is more
complicated, due to the presence of the residual interactions that
give place to configuration mixing. As result, the cross section for
the formation of a state $i $ of the product nucleus is related to
that calculated with DWBA for the formation  from a single-particle
state by
\begin{equation}\Bigl({d\sigma\over d\Omega}\Bigr)_{\hbox{\rm exp}}
= {2J_{B}+1\over2J_{A}+1}{\cal S}_{ij}\Bigl({d\sigma\over
d\Omega}\Bigr)_{\hbox{\rm DWBA}}, \label{dir8}\end{equation} where
the {\it spectroscopic factor} ${\cal S}_{ij} $ measures the weight
of the configuration $j $ used in the DWBA calculation, in the final
state $i $, with the sum-rule
\begin{equation}\sum_j{\cal S}_{ij}=n_i.\label{dir82}\end{equation}
The sum \ref{dir82} embraces all the nucleons in the configurations
$j $ of the product nucleus. The statistical weight
${(2J_{B}+1)/(2J_{A}+1)} $ that appears in the DWBA calculation
involving the angular momentum of the target nucleus {$J_A $} and
final nucleus {$J_{B} $}, is explicitly given in \ref{dir8}.

\section{Heavy ion reactions}

\subsection{Types and properties}

Heavy ion reactions (with $A>4 $) can be
separated into 3 major categories.

1) Due to their large charge, two heavy nuclei feel a strong mutual
Coulomb repulsion. To produce a nuclear reaction the projectile
needs enough energy to overcome the Coulomb barrier. For a very
heavy target, as $^ {238}$U, it is necessary about 5 MeV per
nucleon. Then the wavelength of the projectile is small compared
with the dimensions of the nuclei and classical and semi-classical
methods become useful in the description of the reaction.

2) The projectile carries a large amount of angular momentum and a
good part of it can be transferred to the target in the reaction.
Rotational bands with several dozens of units of angular momentum
can be created. In fact, heavy ion reactions are the best suited to
feed high spin levels.

3) Direct reactions and formation of compound nucleus are also
common processes  in reactions with heavy ions. But some
peculiarities of these are not found in reactions with projectile
nucleons. One of these processes can be understood as intermediate
between a direct reaction and the formation of a compound nucleus.
Fusion does not occur but projectile and target pass a relatively
long time under the mutual action of the nuclear forces. Nuclear
matter is exchanged between both and there is a strong heating of
the two nuclei, with a large transfer of kinetic energy to the
internal degrees of freedom. These are the {\it deep inelastic
collisions}.

The kind of process that prevails depends upon the distance of
closest approach $d $ between the projectile and target. If this
distance is sufficiently large only the long range Coulomb
interaction acts and, for a classical hyperbolic trajectory, $d $ is
related to the impact parameter $b $ and to the energy $E $ of the
projectile by $d={a/2}+\Bigl[({a/2})^2+b^2\Bigr]^{1/2}$ where $a $
is the distance of closest approach in a head-on collision. It is
this is related to $E $ by $a={Z_1Z_2e^2/4\pi\epsilon_0E}.$

Experimentally, the variable under control is the energy $E $ of the
projectile and, for $E $ sufficiently large, $d $  can be small
enough to enter the range of nuclear forces. Collisions near this
limit are called {\it grazing} collisions and  are characterized by
values of $b_{graz} $ and $d_{graz} $. Assuming that there is always
reaction when $b< b_{graz} $, the reaction cross section $\sigma_r $
can be determined geometrically by $\sigma_r=\pi b_{graz}^2.$ The
experimental determination of $\sigma_r $ allows to establish the
value  $d_{ras}=0.5+1.36(A_1^{1/3}+A_2^{1/3}) $  showing that the
distance of grazing collision is somewhat larger than that deduced
from two touching spheres $(1.36\hbox { fm}> r_0=1.2 $ fm).

When the impact parameter is close to $b_{graz} $ one expects
nuclear reactions of short duration, without the contribution of the
compound nucleus  formation. Such reactions are elastic and
inelastic scattering and transfer of few nucleons. When the incident
energy is sufficiently high, small values of $b $ can lead to the
projectile penetrating the target. Depending on the energy and on
the involved masses, the reaction can end in one of the processes
below:

  {\it a) Fusion} - is the preferred process when one has light nuclei
and low energy. There is the formation of a  highly excited compound
nucleus that decays by evaporation of particles and
$\gamma$-radiation emission, leading to a cold residual nucleus. If
the energy in the CM is close to the Coulomb barrier energy the
cross section of compound  nucleus formation starting from two
nuclei is practically equal to the reaction cross  section.

  {\it b) Fission} - When the compound nucleus is heavy the fission
process competes strongly with the evaporation of particles in each
stage of the evaporation process. A very heavy compound nucleus with
a large excitation energy  has a very small probability of arriving
to a cold residual nucleus without fission at some stage of the
de-excitation. The role of the angular momentum  $l $ transmitted to
the target nucleus is also essential. The fission barrier decreases
with the increase of $l $ and for a critical value  $l_{\hbox{\small
crit}} $ the barrier ceases to exist.  A nucleus with angular
momentum greater than $l_{\hbox{\small crit}} $ suffers immediate
fission and this is also a limiting factor in the production of
superheavy elements.

  {\it c) Deep inelastic collision (DIC)} - is a phenomenon
characteristic of reactions involving very heavy nuclei ($A\ > \ 40
$) and with an incident energy of 1 MeV to 3 MeV above the Coulomb
barrier. In DIC the projectile and the target spend some time under
mutual action, exchanging masses and energy but without arriving to
the formation of a compound nucleus. The projectile escapes after
transferring part of its energy and angular momentum to its internal
degrees of freedom  and to the target, with values reaching 100~MeV
and $50\hbar $.

\begin{figure}
[t]
\begin{center}
\includegraphics[
natheight=1.2in, natwidth=1.1in, height=1.5in, width=3.3in
]%
{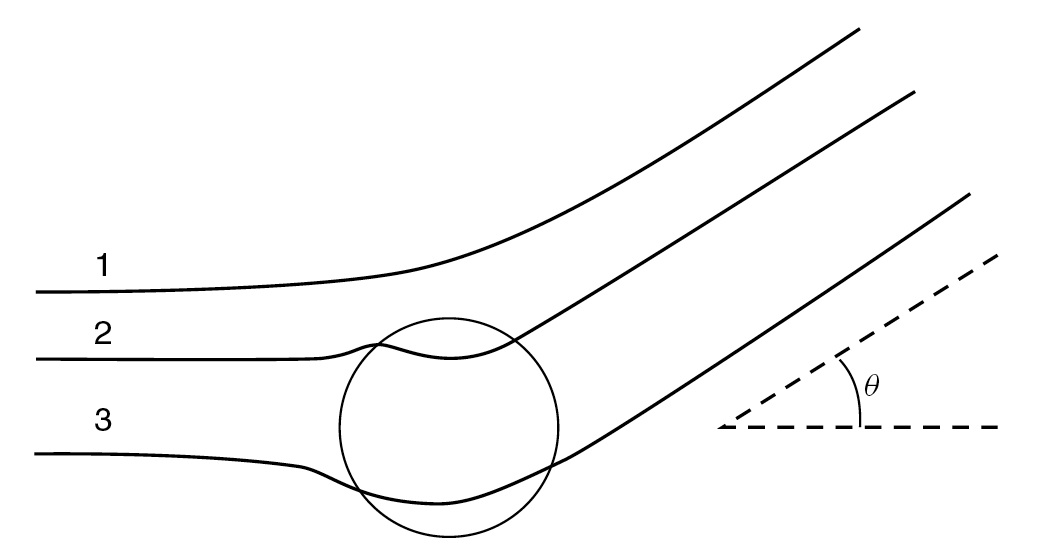}%

\medskip

\includegraphics[
natheight=1.2in, natwidth=1.2in, height=1.7in, width=3.3in
]%
{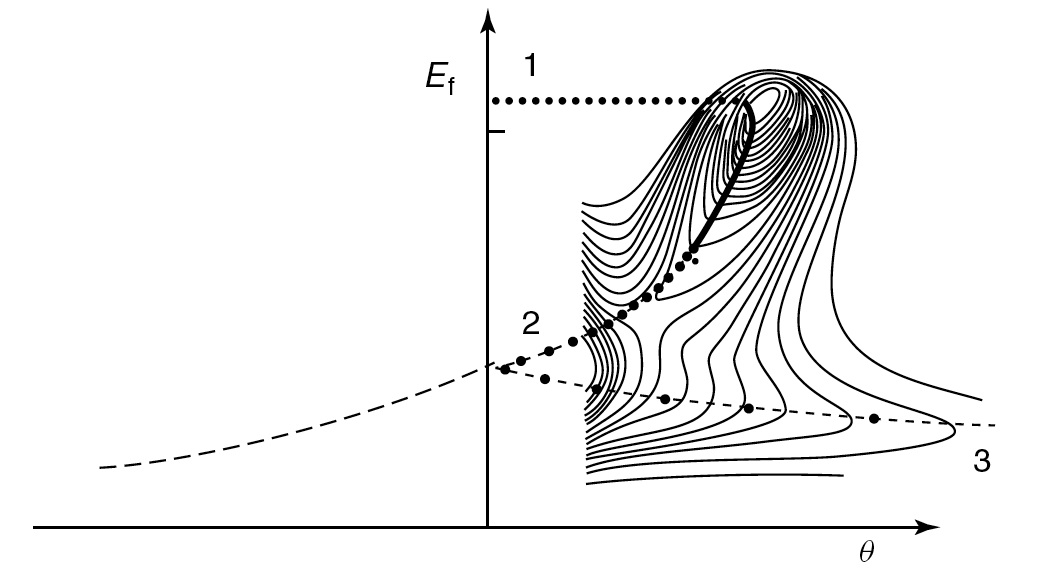}%
\caption{(Top) Reactions with different impact parameters leading to
the same scattering angle $\theta $. (Bottom) Lines of maxima in
topographical diagrams of the final energy against the scattering
angle.}
\label{wilz}%
\end{center}
\end{figure}

One of the  most interesting aspects  of DIC is the correlation
between the energy dissipated in the collision and the scattering
angle in the center of mass. Let us look at figure \ref{wilz}(top).
The trajectory 1 shows the projectile with an impact parameter that
leads it out of the range of nuclear forces. The Coulomb scattering
angle will become larger as the impact parameter decreases. In a
graph as in figure \ref{wilz}(bottom), where one plots the final
energy against the scattering angle, trajectories of type 1 are
located in the upper branch, where there is no dissipation and the
initial kinetic energy stays unaffected. This upper branch has a
maximum value for the scattering angle. At a smaller impact
parameter  the nuclear attractive force begins to act and, with that
the effects of dissipation of DIC. A given Coulomb scattering angle
$\theta $  can also be reached by the combination of the nuclear and
Coulomb forces. Only that now there is loss of energy and the events
are located in the branch 2 of figure \ref{wilz}(b). There still is
no one-to-one correspondence between angle and energy because an
infinite number of trajectories can lead to the same angle $\theta
$. The branch 2 should be understood as a line of maxima in a
three-dimensional representation (called {\it Wilczynski diagram})
where the axis perpendicular to the paper is proportional to the
cross section $d^2\sigma/dEd\theta $. The same angle $\theta $ can
also be obtained by the trajectory 3, with a longer interaction time
between the nuclei and a larger dissipation. As now the projectile
is deflected towards the nucleus, the scattering angle would be
formally $- \theta $ but, as there is no experimental distinction
between $\theta $ and $- \theta $, these events appear as an
independent ramification in the lower part  of the diagram.

\subsection{Superheavy elements} The heaviest element
found in the nature is  $^ {238}_ {\ 92}$U. It is radioactive, but
it survived since its formation in supernovae explosions  because it
has a decay  half-life  of the order of the age of the Earth.
Elements with larger atomic number (transuranic) have shorter
half-lives and have disappeared. They are created artificially
through nuclear reactions using heavy elements as target. Initially,
the projectiles used were light particles: protons, deuterons,
$\alpha $-particles and neutrons. The use of neutron is justified
because the $\beta ^ -$-emission of the compound nucleus increases
the value of $Z $ and it was in this way \cite{MA40} that the first
transuranic element, the neptunium, was obtained: $\hbox{n}+ \ ^
{238}\hbox{U}\rightarrow \ ^ {239}\hbox{U}\rightarrow \ ^
{239}\hbox{Np}+\beta ^ -. $

Reactions with light particles can produce isotopes up to
mendelevium ($Z=101 $), but it is not possible to go beyond that;
the half-lives for $\alpha $-emission or spontaneous fission become
extremely short, turning impracticable the preparation of a target.
The alternative is to place a heavy element under the flux of very
intense neutrons. This can be done using special reactors or using
the rest material of nuclear explosions. The elements einsteinium
($Z=99 $) and fermium ($Z=100 $) were discovered in this way in 1955
but the increasing competition that the  beta decay has with alpha
decay and with spontaneous fission prevents this method to be used
for larger $Z $.

Starting in 1955 heavy ion accelerators began to deliver beams with
high enough intensity and energy to compete in the production of
transuranic isotopes. The first positive result was the production
of two californium isotopes ($Z=98 $) in the fusion  of carbon and
uranium nuclei:
\begin{eqnarray}
_ {\ 6}^{12}\hbox{C} \ + \ _ {\
92}^{238}\hbox{U} \ \rightarrow \ _ {\ 98}^{244}\hbox{Cf}+\hbox{6n},
\nonumber \\
_ {\ 6}^{12}\hbox{C} \ +_ {\ 92}^{238}\hbox{U} \ \rightarrow \ _ {\
98}^{246}\hbox{Cf}+\hbox{4n}. \end{eqnarray}

This  opened the possibility of reaching directly the nucleus one
wants to create from the fusion of two smaller nuclei. The
difficulty of such task is that the cross sections for the
production of heavy isotopes are extremely low. As example, the
reaction $^ {50}_{22}$Ti + $^ {208}_ {\ 82}$Pb $\rightarrow \ ^
{257}_{104}$Rf + n, which produces the element rutherfordium,  has a
cross section of only 5 nb. A small increase in the charges reduces
drastically this value: the cross section for the fusion reaction $^
{58}_{26}$Fe + $^ {208}_ {\ 82}$Pb $\rightarrow \ ^ {265}_{108}$Hs +
n is 4 pb. As comparison, the typical cross sections of DIC for
heavy nuclei are in the range 1-2 b.

In spite of the experimental refinement that these low cross
sections demand, one is able to produce isotopes with charge as
heavy as $Z=118$. The understanding of the mechanisms that lead to
fusion is, however, not fully understood. According to the
traditionally accepted model, the fusion of two nuclei proceeds in
two stages: the formation of a compound nucleus and the
de-excitation of the compound nucleus by evaporation of particles,
preferentially neutrons. The difficulties for the materialization of
the process in very heavy nuclei reside in both stages \cite{Oga06}.

\section{Electromagnetic probes}
\subsection{Coulomb Excitation}

Coulomb excitation is a inelastic scattering process in which a
nucleus excites another nucleus with its electromagnetic field $V$.
This field can be decomposed in terms of a series of multipoles,
e.g. $E1$, $E2$, $M1$, $\cdots$, which carry well defined angular
momenta and parities. At low bombarding energies  $E2$ (quadrupole)
excitations are more common, while at higher energies $E1$
excitations dominate. As an example of a low energy reaction, let us
consider the excitation of a quadrupole state in a head-on collision
below the Coulomb barrier, i.e. for a situation in which the
projectile deccelerates as it approximates the target and stops
before reaching the range of the nuclear interaction, reaccelerating
backwards after that. The differential cross section is given by the
product of the Rutherford differential cross section at 180$^\circ $
and the probability of transition of the target from state $i$ to
state $f$, along the trajectory, measured by the square of
\begin{equation}{d\sigma\over
d\Omega}\Bigr\vert_{\theta=180^\circ}={d\sigma_R\over
d\Omega}\Bigr\vert_{\theta=180^\circ}\times\vert
a_{if}\vert^2.\label{cou1}\end{equation}

The square of $a_{if} $ measures the transition probability from i
to f and this probability should be integrated along the trajectory.
A simple calculation for $a_{if}$ can be done in the case of the
excitation from the ground state $J=0 $ of a deformed nucleus to an
excited state with $J=2 $. The perturbation $V$ comes, in this case,
from the interaction of the projectile charge $Z_Pe$ with the
quadrupole moment of the target nucleus. This quadrupole moment
works as an operator that acts between the initial and final states,
i.e. \begin{equation}V={1\over2}{Z_Pe^2Q_{if}\over
r^3},\label{coulf}\end{equation} where $r$ is the projectile-target
separation distance and
\begin{equation}Q_{if}=\sum_i\int\Psi^*_f(3z_i^2-r_i^2)\Psi_i\,d\tau,\label{coulf2}\end{equation}
where the sum extends to all  protons at the positions $r_i = (x_i,
y_i, z_i)$ in the target. The amplitude is then given by
\begin{equation}a_{if}={4Q_{if}E^2\over3Z_Pe^2\hbar v_0Z_T^2}.\label{aif}
\end{equation}

Using eq. \ref{cou1}for the Rutherford differential cross section at
$\theta=180^\circ $, we obtain
\begin{equation}{d\sigma\over
d\Omega}\Bigr\vert_{\theta=180^\circ}={mE\vert
Q_{if}\vert^2\over18\hbar^2Z_T^2},\label{cou3}\end{equation} which
is an expression that is independent of the charge of the
projectile. It is, on the other hand, proportional to the mass of
the projectile, indicating that heavy ions are more effective for
Coulomb excitation.

The quadrupole moment operator $Q_{if} $ uses the wavefunctions
$\Psi_i $ and $\Psi_f $ of the initial and final states. If those
two wavefunctions are similar, as is the case of an excitation to
the first level of a rotational band, the operator $Q_{if} $ can be
replaced by the  intrinsic quadrupole moment  $Q $. The expression
translates, in this way, the possibility to evaluate the quadrupole
moment from a measurement of the cross section. This has been indeed
a major spectroscopic tool for determining quadrupole momenta (and
transitions) along the nuclear chart.

\subsection{Photonuclear reactions and giant resonances} A photonuclear
reaction is a reaction resulting from the interaction of the
electromagnetic radiation with a nucleus. Therefore, one can access
information which are complementary to Coulomb excitation (and
vice-versa).

\begin{figure}
[t]
\begin{center}
\includegraphics[
natheight=1.2in, natwidth=1.2in, height=2.in, width=3.3in
]%
{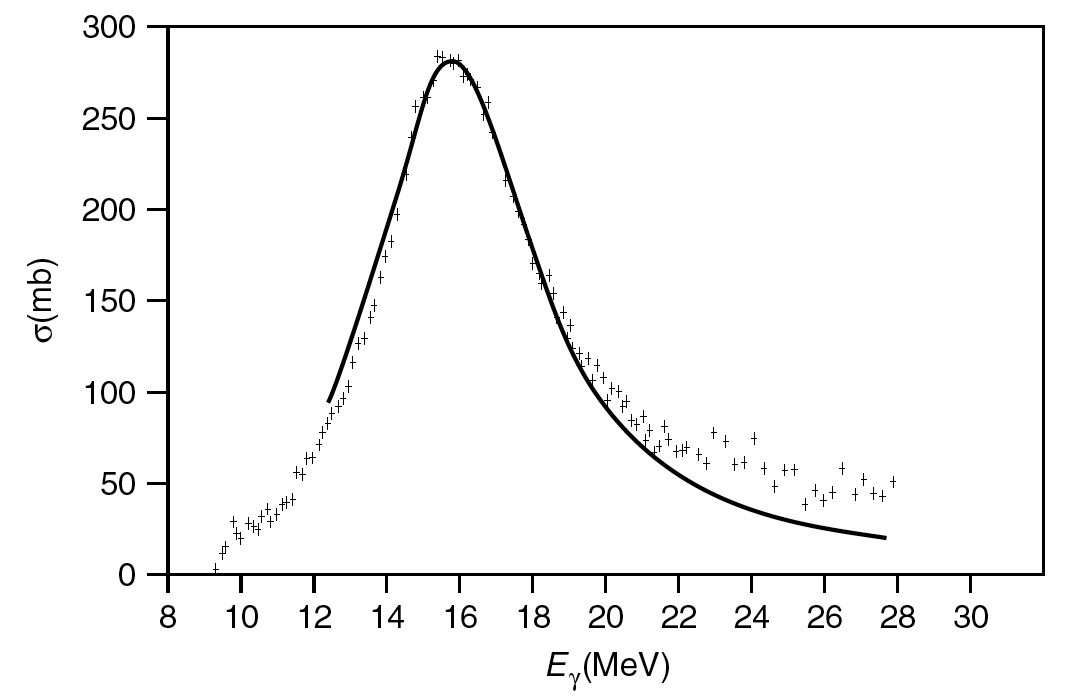}%
\caption{Giant resonance in the absorption of photons by $^{120}$Sn
\cite{Le74}.}
\label{gdr}%
\end{center}
\end{figure}

When the energy of the photon is located above the separation energy
of a nucleon, the cross section of photo-absorption reveals the
presence of characteristic sharp resonances. But when the incident
energy reaches the range of 15-25 MeV, a new behavior appears in the
cross section, with the presence of a wide and large peak, called
{\it   giant electric dipole resonance} ($E1$ excitation).  Figure
\ref{gdr} exhibits the excitation function of photoabsorption of $^
{120}$Sn at photon energies around the electric dipole giant
resonance at 15 MeV.

The giant resonance occurs in nuclei along the whole periodic table,
with the resonance energy decreasing was $E_{GDR} \simeq 80/A^{1/3}$
for $A>20 $.  Their widths are almost all in the range between 3.5
MeV and 5 MeV. In few cases they can reach 7~MeV. They are a
collective excitation, i.e., an excitation involving many nucleons
at once. The time-varying electric field of the photon is very
effective in inducing collective oscillations of protons against
neutrons.

The giant  electric dipole (GDR)  resonance arises from an
excitation that transfers by one unit of angular momentum to the
nucleus ($\Delta l=1 $). If the nucleus is even-even it is taken to
a $1 ^ - $ state. The transition also changes the isospin of 1 unit
($\Delta T=1 $) and, due to that, it is also called {\it isovector
resonance}. The photon can excite less effectively, but still with
appreciable cross sections, giant isoscalar resonances, with $\Delta
T=0 $. But electric quadrupole ($\Delta l=2 $) and electric monopole
($\Delta l=0 $) are observed mostly in reactions with charged
particles. In a giant electric quadrupole resonance the nucleus
oscillates between an spherical (assuming  that this is the form of
the ground state) to an ellipsoidal form. If protons and neutrons
vibrate in phase, we have an isoscalar resonance ($\Delta T=0 $) and
if they are oscillate in opposite phase we have an isovector
resonance ($\Delta T=1 $). The giant monopole resonance is a very
special case of nuclear excitation where the nucleus contracts and
expands radially, maintaining its original form but changing its
volume. It is also called  {\it breathing mode}. It can also happen
in isoscalar and isovector  forms. It is an important way to study
the {\it compressibility of  nuclear matter}.

Besides electric giant resonances, associated to a variation in the
form of the nucleus, magnetic giant resonances exist, involving {\it
spin vibrations}. In these, nucleons with spin up move out of phase
with nucleons with spin down. The number of nucleons involved in the
process cannot be very large because it is limited by the  Pauli
principle. The magnetic resonances can also separate into isoscalar
resonances, where protons and neutrons of  same spin vibrate against
protons and neutrons of opposite spin, and isovector, where protons
with  spin up and neutrons with  spin down vibrate against their
corresponding partners with opposite spins. The last cases, are
effectively probed in charge-exchange reactions, i.e. when the
projectile charge changes to Z-1 or Z+1, as in (d,p) and (d,n). They
are known as {\it giant Gamow-Teller resonances}.

Giant resonances can also be produced in an excited nucleus, a case
known as the Brink-Axel hypothesis \cite{Ax62}. A special case is
when  two giant resonances are excited in the same nucleus.  The double giant dipole resonance (or {\it multiphonon
giant resonance} was observed for the first time in reactions with
double charge exchange induced by pions in the $^ {32}$S
\cite{Mo88}. But Coulomb excitation proved to be their best probe,
as shown theoretically in refs. \cite{BB88,Sc93}.

\subsection{Electron scattering}
Electron scattering experiments have provided a rich database on
most stable or long-lived nuclei. Electrons are structureless
point-like objects that only interact electromagnetically.
Therefore, electron scattering avoids the complexity of the strong
interaction between the projectile and the target, and provides
clean information about the charge distribution in the nucleus.

Considering relativity and the spin of the electrons, the
differential cross section for elastic scattering by spinless
point-like nuclei can be expressed by the {\it Mott scattering
formula} (for simplicity, we use $\hbar=c=1$ units):
\begin{equation}
\left(  \frac{d\sigma}{d\Omega}\right)
_{Mott}=\frac{Z_{T}^{2}e^{4}\cos ^{2}\left(  \theta/2\right)
}{4p_{0}^{2}\sin^{4}\left(  \theta/2\right) \left[  1+\left(
2p_{0}/M\right)  \sin^{2}\left(  \theta/2\right)  \right] }
\end{equation}
where $Z_Te$ and $M$ are the charge and mass of the target nucleus
respectively, and $p_{0}$ is the momentum of the incoming electron.
Since the nucleus is not a point-like particle, the formula is
modified by adding the nuclear electric and magnetic form factors,
which contain information of the charge and magnetization
distributions inside the nucleus.

The \textit{Rosenbluth formula} accounts for the electron spin and
explicitly expresses the
cross section for arbitrary nuclei as%
\begin{align*}
\left(  \frac{d\sigma}{d\Omega}\right)    & =\left(
\frac{d\sigma}{d\Omega
}\right)  _{Mott}\left\{  A_{0}\left(  q^{2}\right)  +\right.  \\
& \left.  B_{0}\left(  q^{2}\right)  \left[  \frac{1}{2}+\left(
1+\frac {q^{2}}{4M^{2}}\right)  \right]  \tan^{2}\left(
\theta/2\right)  \right\}
,\newline%
\end{align*}
where $q^{2}=\left(  \mathbf{p}_{f}-\mathbf{p}_{i}\right)
^{2}-\left( E_{f}-E_{i}\right)  ^{2}$ is the 4-momentum transfer
squared, where $\mathbf{p}_{i}\left(  \mathbf{p}_{f}\right)  $ and
$E_{i}\left( E_{f}\right)  $ are the electron incoming (outgoing)
momentum and energy. $A_{0}(q^{2})$ and $B_{0}(q^{2})$, are
functions of $q^{2}$, are the form factors associated with the
charge and magnetization distribution of the nucleus respectively.
For a spin-0 nucleus:
\begin{equation}
A_{0}(q^{2})=\frac{G_{E}^{2}(q^{2})}{1+q^{2}/4M^{2}}\ \ \ \ \ \text{and}%
\ \ \ \ \ B_{0}(q^{2})=0,
\end{equation}
where $G_{E}(q^{2})$ is the {\it Sachs charge form factor}, and the
factor $1+q^{2}/4M^{2}$ is the kinematical recoil correction. For
$q\rightarrow0$, $q^{2}\approx \mathbf{q}^{2}$, where \textbf{q} is
the three-momentum transfer, and to leading order in powers of
\textbf{q}:
\begin{equation}
G_{E}(q^{2})\simeq
G_{E}(\mathbf{q}^{2})=\int\rho_{T}(\mathbf{r})e^{i\mathbf{q\cdot
r}}d^{3}r.
\end{equation}

One thus sees that electron scattering is related to the Fourier
transform of the nuclear charge density, $\rho_{T}(\mathbf{r})$. If
the function $G_{E}({\bf q}^{2})$ is mapped for a sufficient number
of momentum transfers ${\bf q}^{2}$ the Fourier transform can be
inverted and $\rho_{T}(\mathbf{r})$ can be mapped with precision.
For heavy elements ($A>10$), this technique shows that the charge
radius is closely proportional to  1.12 $A^{1/3}$ fm, where $A$ is
the nuclear mass number. This indicates the density saturation for
the nuclear matter.

When $\mathbf{q}^2=0$, inelastic electron scattering probes the same
multipolarity transitions ($E1$, $E2$, $\cdots$) as in Coulomb
excitation or with real photons. However, in contrast to the later
probes, electron scattering also provides information on nuclear
excitations for cases in which $\mathbf{q}^2\neq 0$. This is useful
in many aspects. For example, at high momentum transfers, the
momentum transfer $\textbf{q}$ can be absorbed by a nucleon in the
nucleus which acquires an energy of the order of
$\textbf{q}^2/2m_N^*$, thus probing the effective mass $m_N^*$ of a
nucleon in the nucleus. This is of relevance to understand the
effects of nucleon-nucleon interactions in the nuclear environment.
The $q$-region where this occurs is termed by {\it quasi-free
scattering}. As electrons can penetrate the nuclei, they can also
more effectively probe monopole (or breathing mode) excitations of
the nuclei, a case in which the nuclear mass distribution vibrates
radially. This is of importance for determining the {\it
compressibility modulus} of the nuclear matter. Finally, in
high-energy electron scattering the electron can penetrate deeply
inside the nucleons and probe the spin and charge distributions of
quarks and gluons inside the nucleon \cite{Fran05}.

\section{Relativistic nuclear collisions}

\subsection{Transport theories and equation of state}
As the bombarding energy in nucleus-nucleus collisions increases the
structure aspects of the nuclei become less relevant. Except for the
bulk properties of the nuclei (size and number of nucleons), the
physics involved is primarily due to the individual, and sometimes
collective, hadronic collisions. Several theoretical tools are used
to describe these reactions, and in particular we quote (a) {\it
time-dependent Hartree-Fock} (TDHF), (b) {\it anti-symmetrized
molecular dynamics} (AMD), (c) {\it Boltzmann-Uehling-Uhlenbeck}
(BUU}, etc.

At intermediate energies of $E_{lab}\sim100-1000$ MeV/nucleon the
nucleons and the products of their collisions can be described
individually and their propagation can be described by semiclassical
equations. One of such equations, and perhaps the most popular in
such studies, is the so-called \textit{Boltzmann-Uehling-Uhlenbeck}
(BUU) equation:
\begin{align}
&  \frac{\partial f}{\partial t}+\left(
\frac{\mathbf{p}}{m}+\mathbf{\nabla }_{\mathbf{p}}U\right)
\cdot\mathbf{\nabla}_{\mathbf{r}}f-\mathbf{\nabla
}_{\mathbf{r}}U\cdot\mathbf{\nabla}_{\mathbf{r}}f=\nonumber\\
&  \int d^{3}p_{2}\int d\Omega\;\sigma_{NN}\left(  \Omega\right)
\left\vert
\mathbf{v}_{1}-\mathbf{v}_{2}\right\vert \nonumber\\
&  \times\left\{  f_{1}^{\prime}f_{2}^{\prime}\left[  1-f_{1}\right]
\left[ 1-f_{2}\right]  -f_{1}f_{2}\left[  1-f_{1}^{\prime}\right]
\left[
1-f_{2}^{\prime}\right]  \right\}  , \label{BE}%
\end{align}

If $dN$\ is the number of particles in the volume element $d^{3}r$\
and whose momenta fall in the momentum element $d^{3}p$\ at time
$t$, then the distribution \ function{\small \ }$f\left(
\mathbf{r,p},t\right)  $\ is given by $
dN=f\left(  \mathbf{r,p},t\right)  d^{3}rd^{3}p \label{distbf1}%
$. Thus the BUU equation is an equation for the distribution
function $f\left(  \mathbf{r,p},t\right)$. To account for the effect
of each particle interacting with all others, one
introduces the concept of \textit{mean-field}, $U\left(  \mathbf{r,p}%
,t\right)  .$\ This mean-field exerts a force on each particle,
given by{\small \ }$-\nabla_{\mathbf{r}}U\left(
\mathbf{r,p},t\right) .$ Also, the momentum dependence of the
potential introduces a dependence through the derivative
$-\nabla_{\mathbf{p}}U\left( \mathbf{r,p},t\right) .$

Due to the nucleon-nucleon collisions, the distribution function
within $d^3rd^3p$ can also be modified by nucleons leaving (or
entering) this volume. This is taken care by the {\it collision
term}, i.e. the right-hand-side of the BUU equation. $\sigma_{NN}$
is the nucleon-nucleon differential scattering cross section, ${\bf
v}_1$ and ${\bf v}_2$ are the velocities of two colliding nucleons.
The first factor inside braces are for collisions populating the
volume element and the second term for those depleting it. The
factors $(1-f)$ account for Pauli blocking of final occupied states.
The integrals average over scattering angle and over all collisions
within $d^3rd^3p$. The BUU equation falls in the category of what
one calls \textit{quantum transport theories}. Hadronic transport
theories have been quite successful in applications, describing a
multitude of measured particle spectra.

\begin{figure}[t]
\begin{center}
\includegraphics[
natheight=38.402901in, natwidth=32.361301in, height=4.2in,
width=3.25in ]{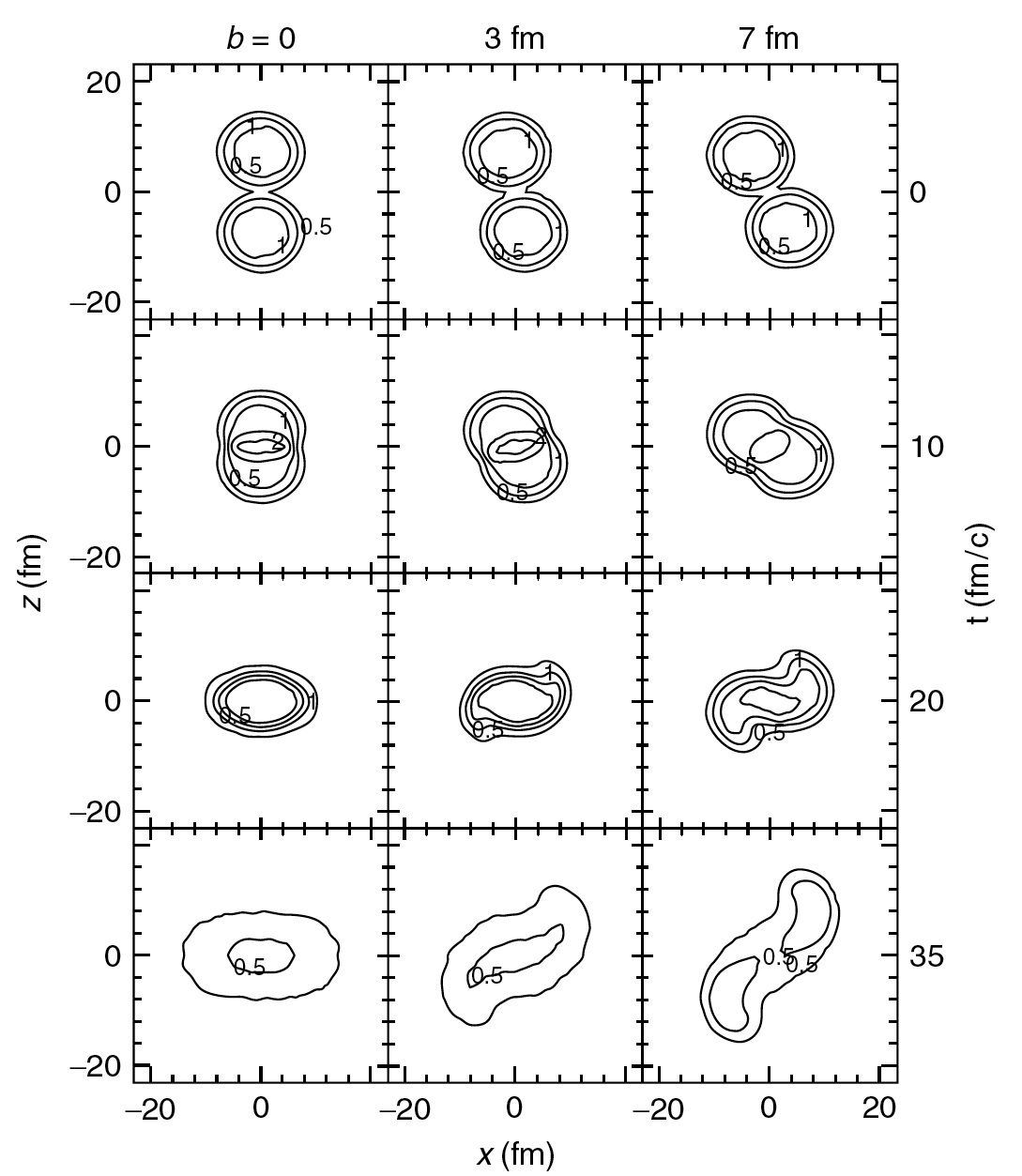}
\end{center}
\caption{- Contour plots of baryon density in the reaction plane in
Au + Au collisions at 400 MeV/nucleon. The displayed contour lines
are for the
densities $\rho/\rho_{0}$ = 0.1, 0.5, 1, 1.5, and~2 \cite{Da95}. }%
\label{contours}%
\end{figure}

Eq. (\ref{BE}) needs as basic ingredients the mean field $U$\ and
the cross section $\sigma_{NN}$. Because these two quantities are
related to each other, one should in principle derive them in a
self-consistent microscopic approach, as in the {\it Brueckner
theory}. However, in practice the simulations are often done with a
phenomenological mean field and free nuclear cross sections.

An important ingredient in the transport theory calculations is the
compressibility $K$ of nuclear matter, which refers to the second
derivative
of the compressional energy $E$ with respect to the density:%
\begin{equation}
K=9\rho^{2}\frac{\partial^{2}}{\partial\rho^{2}}\left(
\frac{E}{A}\right)
\ . \label{Kcomp}%
\end{equation}
This is an important quantity, e.g., for nuclear astrophysics.
Supernova models might or not lead to explosions depending on the
value of $K$. The central collisions of heavy nuclei are one of the
few probes of this quantity in the laboratory. The dependence of the
calculations on $K$ follow from the dependence of the mean field
potential $U$ ($U\sim E/A+$ kinetic energy terms) on the particle
density $\rho$. A typical
parametrization for $U$ is the Skyrme parametrization%
$ U=a{\rho}/{\rho_{0}}+b\left(  {\rho}/{\rho_{0}}\right) ^{\sigma} .
$

The output of eq. (\ref{BE}) is the distribution function{\small \ }%
$f(\mathbf{r,p},t)$, which allows one to calculate many properties
of heavy-ion collisions. Let us quote \textit{collective flows},
proton and neutron production rates, (sub-threshold and above
threshold) pion and kaon yields, etc. Combining eq. (\ref{BE}) with
a {\it phase-space coalescence model}, one can also calculate such
quantities as exclusive flows and intermediate fragment formations.

The dynamics of the central high-energy reactions can be broken down
into several stages. Baryon-density contour-plots are shown in
fig.~\ref{contours} for 400 MeV/nucleon Au+Au collisions at $b=0$,
which will serve to illustrate our points.

Following an initial interpenetration of projectile and target
densities, the $NN$ collisions begin to thermalize matter in the
overlap region making the momentum distribution there centered at
zero momentum in the CM. The density in the overlap region rises
above normal and a disk of excited and compressed matter forms at
the center of the system. More and more matter dives into the region
with compressed matter that begins to expand in transverse
directions. At late stages, when the whole matter is excited,
transverse expansion predominates.

\subsection{Kinematics}

In relativistic nucleus-nucleus collisions other definitions are
best suited for discussing relations between energies, momenta and
angles. The \textit{rapidity} is a variable frequently used to
describe the behavior of
particles in inclusively measured reactions. It is defined by%
\begin{equation}
y=\frac{1}{2}\ln\left(
\frac{E+p_{\parallel}}{E-p_{\parallel}}\right)
\label{rapidity}%
\end{equation}
which corresponds to $ \tanh y=\frac{p_{\parallel}}{E},$ where $y$
is the rapidity, $p_{\parallel}$\ is the longitudinal momentum along
the direction of the incident particle, $E$ is the energy, both
defined for a given particle. The accessible range of rapidities for
a given reaction is determined by the available center-of-mass
energy and all participating particles' rest masses. One usually
gives the limit for the incident particle,
elastically scattered at zero angle:%
\begin{equation}
\left\vert y_{\max}\right\vert =\ln\left[  \frac{E+p}{m}\right]
=\ln\left( \gamma+\gamma\beta\right)
\end{equation}
where $\beta\equiv v$ is the velocity and all variables referring to
the through-going particle given in the desired frame of reference
(e.g. in the center of mass).

Note that $\partial y/\partial p_{\parallel}=1/E.$ A Lorentz boost
$\beta $\ along the direction of the incident particle adds a
constant, $\ln\left( \gamma+\gamma\beta\right)  $, to the rapidity.
Rapidity differences, therefore, are invariant to a Lorentz boost.
Statistical particle distributions are flat in $y$ for many physics
production models. Frequently, the simpler variable
\textit{pseudorapidity} $\eta$ is used instead of rapidity (and
sloppy language mixes up the two variables).

The pseudorapidity is a handy variable to approximate the rapidity
if the mass
and momentum of a particle are not known. It is an angular variable defined by%
\begin{equation}
\eta=-\ln\left[  \tan\left(  \frac{\theta}{2}\right)  \right]
\label{pseudorapidity}%
\end{equation}
whose inverse function is%
$\theta=2\arctan\left(  e^{-\eta}\right)$, where $\theta$ is the
angle between the particle being considered and the undeflected
beam. $\eta$ is the same as the rapidity \textit{y} if one sets
$\beta=1$ (or $m=0$). Statistical distributions plotted in
pseudorapidity rather than rapidity undergo transformations that
have to be estimated by using a kinematic model for the interaction.

\subsection{The quark-gluon plasma (QGP)}

\begin{figure}[ptb]
\begin{center}
\includegraphics[
natheight=6.572600in, natwidth=7.437400in, height=2.8669in,
width=3.2413in ]{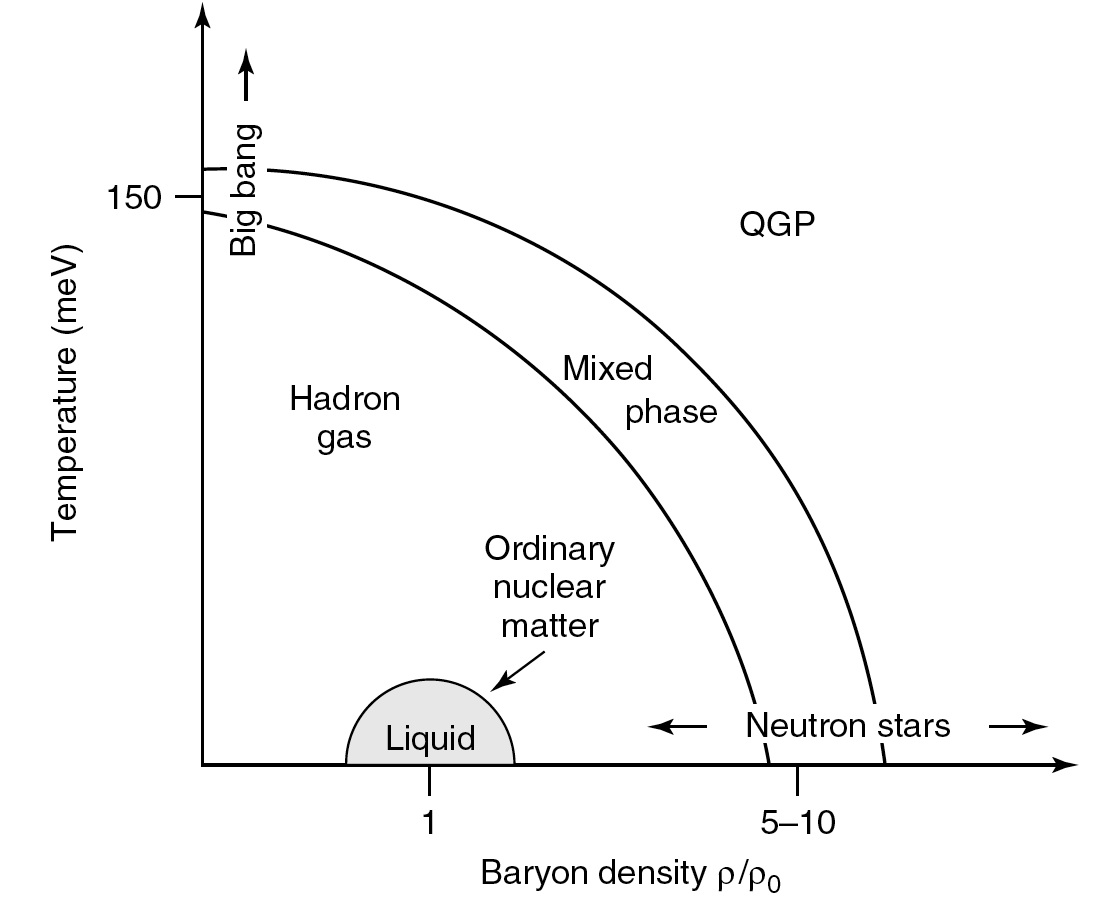}
\end{center}
\caption{- As water comes in different phases (solid, liquid, gas),
so nuclear matter can come in its normal hadronic form \ and at
sufficiently high temperature and density, in the form of a
deconfined \ state of quarks and gluons. The diagram shows how
nuclear matter should behave as a function of
dentistry and temperature.}%
\label{qp}%
\end{figure}

The primary motivation for studying ultra-relativistic heavy ion
collisions is to gain an understanding of the equation of state of
nuclear, hadronic and partonic matter, commonly referred to as
nuclear matter. Displayed in fig. \ref{qp} is a schematic phase
diagram of nuclear matter. The behavior of nuclear matter as a
function of temperature and density (or pressure), shown in fig.
\ref{qp}, is governed by its equation of state.

Conventional nuclear physics is concerned primarily with the lower
left portion of the diagram at low temperatures and near normal
nuclear matter density. Here normal nuclei exist and at low
excitation a liquid-gas phase transition is expected to occur. This
is the focus of experimental studies using low energy heavy ions. At
somewhat higher excitation, nucleons are excited into baryonic
resonance states, along with accompanying particle production and
hadronic resonance formation. In relativistic heavy ion collisions,
such excitation is expected to create hadronic resonance matter.

We now briefly discuss \ the QGP signatures in nucleus-nucleus
collisions. For more details see, e.g., ref. \cite{wong,HW04}. One
group of such signatures can be classified as \textit{thermodynamic
variables.} This class involves determination of the energy density
$\epsilon$, pressure $P$ , and entropy density $s$ of the
interacting system as a function of the temperature $T$ and the
baryochemical potential $\mu_{B}$ . Experimental observables can be
identified with these variables and thus their relative behavior can
be determined. If a phase transition to QGP occurs, a rapid rise in
the effective number of degrees of freedom, expressed by
$\epsilon/T^{4}$ or $s/T^{3}$, should be observed over a small range
of $\ T$. The variables $T$, $s$, and $\epsilon$, can be identified
with the average transverse momentum $\left\langle
p_{T}\right\rangle $, the hadron rapidity density $dN/dy$, and the
transverse energy density $dE_{T}/dy$, respectively. The transverse
energy produced in the interaction is $
E_{T}=\sum_{i}E_{i}\sin\theta_{i}\ , $ where $E_{i}$ and
$\theta_{i}$\ are the kinetic energies of the ejectiles and the
emission angles.
\begin{figure}[ptb]
\begin{center}
\includegraphics[
natheight=6.646100in, natwidth=9.646100in, height=2.2in, width=3.3in
]{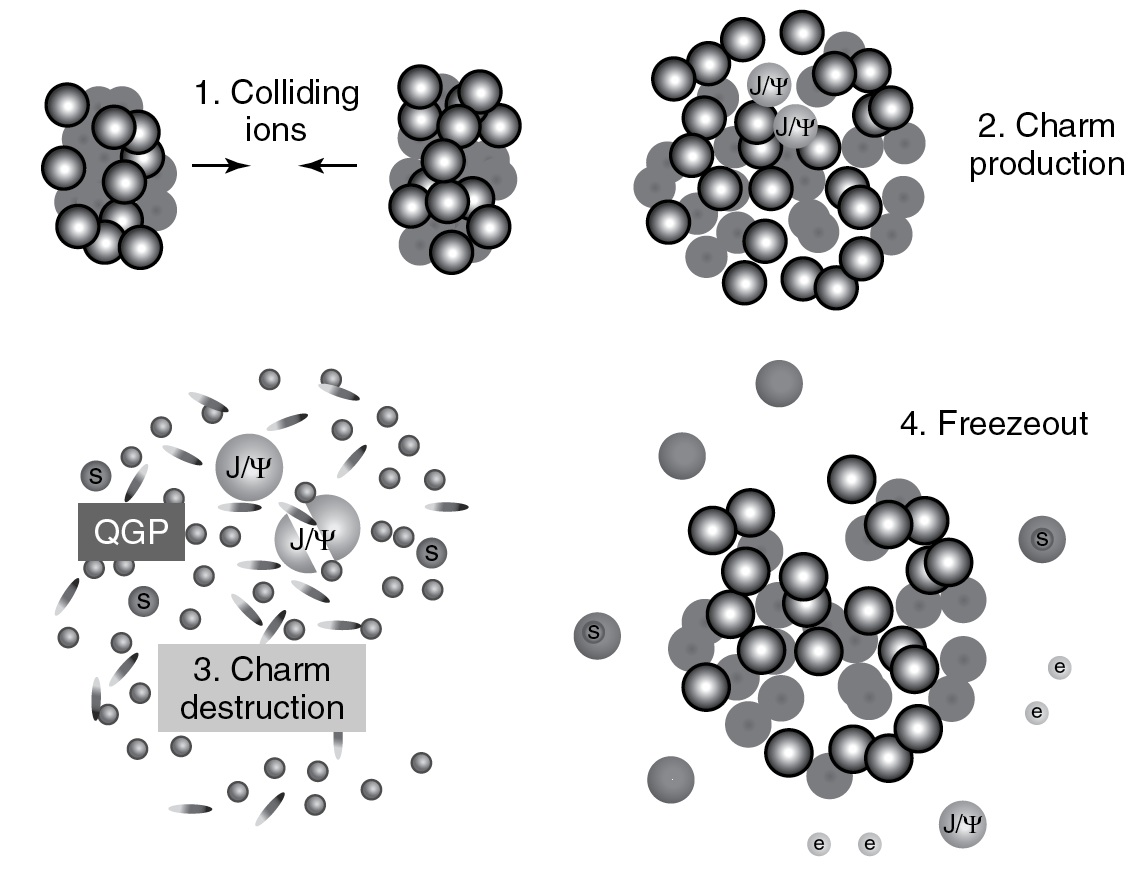}
\end{center}
\caption{- Formation and evolution of J/$\Psi$ particles in
relativistic heavy ion collisions. After the formation, the
J/$\Psi$s \ are dissociated in the plasma due to color screening.
The end effect is a smaller number J/$\Psi$s of
\ than expected from pure hadron-hadron multiple collisions.}%
\label{caca3}%
\end{figure}

Electromagnetic (EM) probes, such as photons and leptons, provide
information on the various stages of the interaction without
modification by final state interactions. These probes may provide a
measure of the thermal radiation from a QGP, if a region of photon
energy, or equivalently lepton pair invariant mass, can be isolated
for emission from a QGP relative to other processes. However, the
yields for EM probes are small relative to background processes,
which are primarily EM decays of hadrons and resonances. Lepton
pairs from the QGP are expected to be identifiable in the 1-10 GeV
invariant mass range. The widths and positions of the $\rho$,
$\omega$, and $\phi$ peaks in the lepton pair invariant mass
spectrum are expected to be sensitive to medium-induced changes of
the hadronic mass spectrum

The production of J/$\Psi$ particles in a quark-gluon plasma is
predicted to be suppressed (see fig. \ref{caca3}). This is a result
of the \textit{Debye screening} of a $c\overline{c}$ pair, initially
formed in the QGP by fusion of two incident gluons. Less tightly
bound excited states of the $c\overline{c}$ system, such as
$\Psi^{\prime}$ and $\chi_{c}$ , are more easily dissociated and
will be suppressed even more than the J/$\Psi$.

A long-standing prediction for a signature of QGP formation is the
enhancement of strange hadrons. The production of strange hadrons
relative to nonstrange hadrons is suppressed in hadronic reactions.
This suppression increases with increasing strangeness content of
the hadron. In a QGP the strange quark content is rapidly saturated
by $s\overline{s}$ pair production in gluon-gluon reactions,
resulting in an enhancement in the production of strange hadrons.
Thus, multi-strange baryons and strange antibaryons are predicted to
be strongly enhanced when a QGP is formed.

The connection between energy loss of a quark and the
color-dielectric polarizability of the medium can be established in
analogy with the theory of electromagnetic energy loss. Although
radiation is a very efficient energy loss mechanism for relativistic
particles, it is strongly suppressed in a dense medium by the
\textit{Landau-Pomeranchuk effect} \cite{wong}. Adding the two
contributions, the stopping power of a quark-gluon plasma is
predicted to be higher than that of hadronic matter. A quark or
\textit{gluon jet} propagating through a dense medium will not only
loose energy but will also be deflected. This effect destroys the
coplanarity of the two jets from a hard parton-parton scattering
with the incident beam axis. The angular deflection of the jets also
results in an azimuthal asymmetry. The presence of a quark-gluon
plasma is also predicted to attenuate the emission of jet pairs
opposite to the trigger jet.

All of the above quark-gluon plasma signatures have been studied to
maturity at the CERN (European Organization for Nuclear Research) - SPS (Super Proton Synchrotron) and the
BNL (Brookhaven National Laboratory) - RHIC (Relativistic Heavy Ion Collider)
facilities. For more details, the following references discuss: (a)
direct photons \cite{Agg00,Bat06}, (b) dileptons \cite{Dam06}, (c)
J/$\Psi$ production \cite{Ada06} and (d) quark in-medium
attenuation, also known as high-$p_T$ quenching or jet attenuation
\cite{Bue06}.

\section{Nuclear reactions in stars}

\subsection{Hydrogen and CNO cycles}

Energy production in stars is a well known process. The initial
energy which ignites the process arises from the gravitational
contraction of a mass of gas. The contraction increases the
pressure, temperature, and density, at the center of the star until
values able to start the thermonuclear reactions, initiating the
star lifetime. The energy liberated in these reactions yield a
pressure in the plasma, which opposes compression due to
gravitation. Thus, an equilibrium is reached for the energy which is
produced, the energy which is
liberated by radiation, the temperature, and the pressure.%
\begin{figure}
[t]
\begin{center}
\includegraphics[
natheight=7.805800in, natwidth=12.903000in, height=2.10in,
width=3.3in
]%
{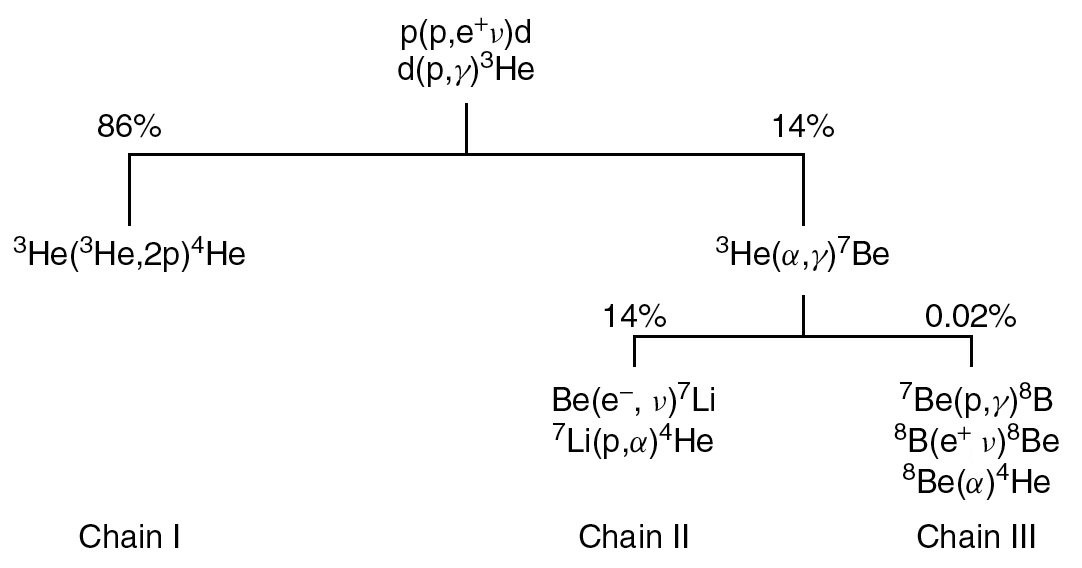}%
\caption{The p-p chain reaction (p-p cycle). The percentage for the
several branches are calculated in the center of the Sun \cite{Bah89}.}%
\label{suncycle}%
\end{center}
\end{figure}

The Sun is a star in its initial phase of evolution. The temperature
in its surface is $6000^{\circ}$ C, while in its interior the
temperature reaches $1.5\times10^{7}$ K, with a pressure given by
$6\times10^{11}$ atm and density 150 g/cm$^{3}$. The present mass of
the Sun is $M_{\odot}=2\times10^{33}$ g and its main composition is
hydrogen (70\%), helium (29\%) and less than 1\% of more heavy
elements, like carbon, oxygen, etc.

What are the nuclear processes which originate the huge
thermonuclear energy of the Sun, and that has last $4.6\times10^{9}$
years (the assumed age of the Sun)? It cannot be the simple fusion
of two protons, or of $\alpha$-particles,
or even the fusion of protons with $\alpha$-particles, since neither $_{2}%
^{2}$He$,$ $_{4}^{8}$Be$,\;$or $_{3}^{5}$Li, are stable. The only
possibility is the proton-proton fusion in the form
\begin{equation}
\mathrm{p}+\mathrm{p}\longrightarrow\mathrm{d}+\mathrm{e}^{+}+\nu
_{e},\label{astroeq1}%
\end{equation}
which occurs via $\beta$-decay, i.e., due to the weak-interaction.
The cross section for this reaction for protons of energy around 1
MeV is very small, of the order of $10^{-23}\;$b. The average
lifetime of protons in the Sun due to the transformation to
deuterons by means of eq. (\ref{astroeq1}) is about 10$^{10}$ y.
This explains why the energy radiated from the Sun is approximately
constant in time, and not an explosive process.

The deuteron produced in the above reaction is consumed almost
immediately in
the process%
\begin{equation}
\mathrm{d}+\mathrm{p}\longrightarrow\ \ _{2}^{3}\mathrm{He}+\gamma
.\label{astroeq2}%
\end{equation}

The resulting $_{2}^{3}$He reacts by means of%
\begin{equation}
\ \ _{2}^{3}\mathrm{He}+\ \ _{2}^{3}\mathrm{He}\longrightarrow\ \ _{2}%
^{4}\mathrm{He}+2\mathrm{p},\label{astroeq3}%
\end{equation}
which produces the stable nucleus $_{2}^{4}$He with a great energy
gain, or by
means of the reaction%
\begin{equation}
\ \ _{2}^{3}\mathrm{He}+\ \ _{2}^{4}\mathrm{He}\longrightarrow\ \ _{4}%
^{7}\mathrm{Be}+\gamma.\label{astroeq4}%
\end{equation}
In the second case, a chain reaction follows as
\begin{equation}
\ \ _{4}^{7}\mathrm{Be}\ +\ \ \mathrm{e}^{-}\longrightarrow\ \ _{3}%
^{7}\mathrm{Li}\ +\ \nu_{e},\;\;\;\;\;\;\ _{3}^{7}\mathrm{Li}+\ \ \mathrm{p}%
\longrightarrow2\left(  _{2}^{4}\mathrm{He}\right)  ,\label{astroeq5}%
\end{equation}
or
\begin{equation}
\ \ _{4}^{7}\mathrm{Be}\ +\ \ \mathrm{p}\longrightarrow\ \ _{5}^{8}%
\mathrm{B}\ +\gamma,\;\;\;\;\;\ _{5}^{8}\mathrm{B}\longrightarrow\ \
2\left(
_{2}^{4}\mathrm{He}\right)  +\mathrm{e}^{+}+\nu_{e}.\label{astroeq6}%
\end{equation}

The chain reaction (\ref{astroeq1})-(\ref{astroeq6}) is called the
\textit{hydrogen cycle.} The result of this cycle is the
transformation of four protons into an $\alpha$-particle, with an
energy gain of 26.7 MeV, about
20\% of which are carried away by the neutrinos (see fig. \ref{suncycle}).%
\begin{figure}
[t]
\begin{center}
\includegraphics[
natheight=6.986000in, natwidth=10.527400in, height=2.2in,
width=3.2in
]%
{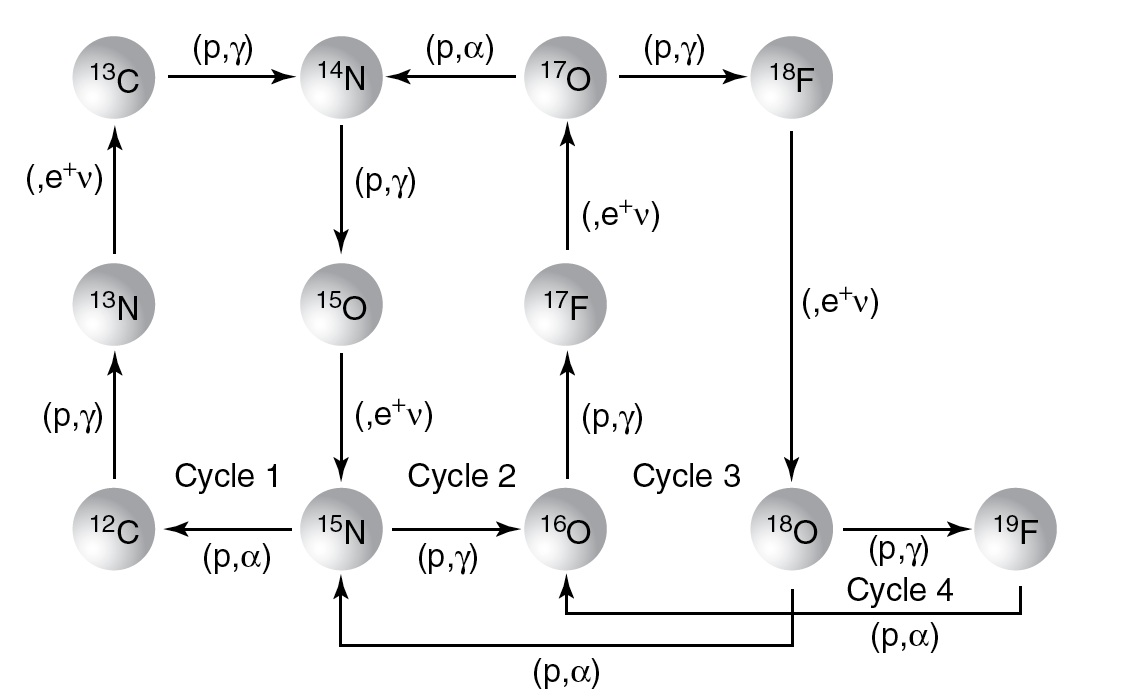}%
\caption{The CNO cycle. (courtesy of Frank Timmes).}%
\label{CNO_1}%
\end{center}
\end{figure}

If the star contains heavier elements, another cycle can occur; the
\textit{carbon cycle}, or \textit{CNO cycle} \cite{Be39}. In this
cycle the carbon, oxygen, and nitrogen nuclei are catalyzers of
nuclear
processes, with the end product also in the form 4p$\longrightarrow$ $_{2}%
^{4}$He. \ fig. \ref{CNO_1} describes the CNO cycle. Due to the
larger Coulomb repulsion between the carbon nuclei, it occurs at
higher temperatures (larger relative energy between the participant
nuclei), up to $1.4\times10^{7}$ K. In the Sun the hydrogen cycle
prevails. But, in stars with larger temperatures the CNO cycle is
more effective.

After the protons are transformed into helium at the center of a
star like our Sun, the fusion reactions start to consume protons at
the surface of the star. At this stage the star starts to become
a\textit{\ red giant.} The energy generated by fusion increases the
temperature and expands the surface of the star. The star luminosity
increases. The red giant contracts again after the hydrogen fuel is
burned.

Other thermonuclear processes start. The first is the helium burning
when the temperature reaches $10^{8}$ K and the density becomes
$10^{6}$ g.cm$^{-3}$. Helium burning starts with the triple capture
reaction
\begin{equation}
3\left(  _{2}^{4}\mathrm{He}\right)  \longrightarrow\ \ _{\ 6}^{12}%
\mathrm{C}+7.65\ \mathrm{MeV},\label{astroeq7}%
\end{equation}
followed by the formation of oxygen via the reaction
\begin{equation}
\ \ _{\ 6}^{12}\mathrm{C}+\ \ _{2}^{4}\mathrm{He}\longrightarrow
\ \ _{\ 8}^{16}\mathrm{O}+\gamma.\label{astroeq8}%
\end{equation}

For a star with the Sun mass, helium burning occurs in about
$10^{7}$ y. For a much heavier star the temperature can reach
$10^{9}$ K$.$ The compression process followed by the burning of
heavier elements can lead to the formation of iron and nickel. After
that the thermonuclear reactions are no more energetic and the star
stops producing nuclear energy.

\subsection{Synthesis of heavier elements}

\begin{figure}
[t]
\begin{center}
\includegraphics[
natheight=20.347300in, natwidth=20.694099in, height=3.in, width=3.in
]%
{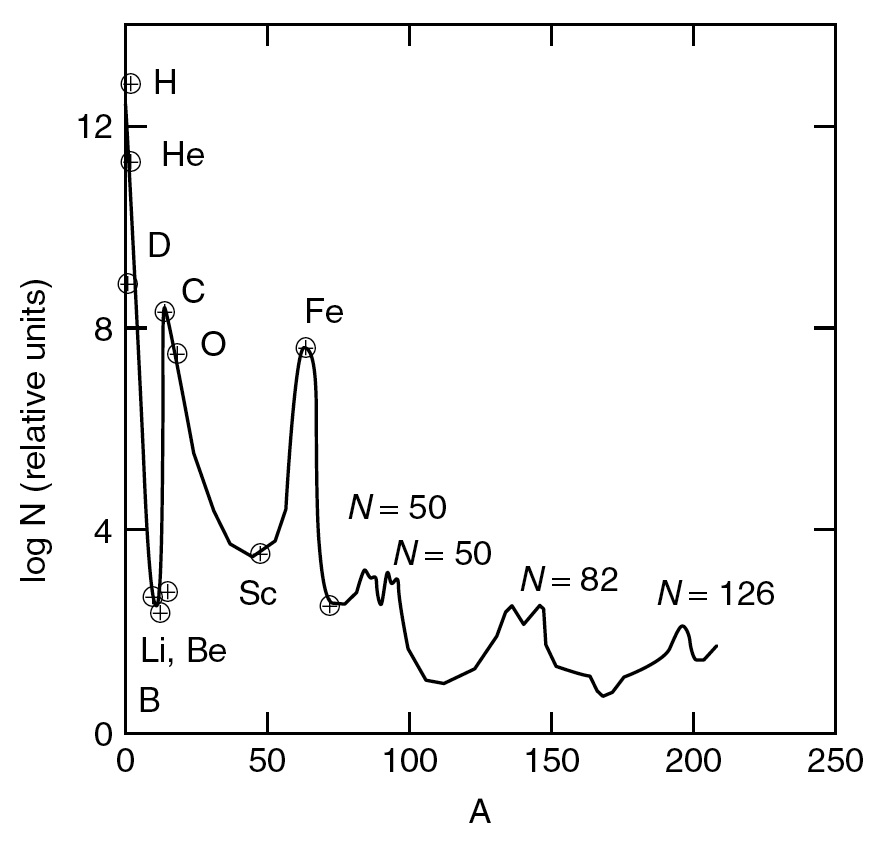}%
\caption{Relative distribution of elements in our galaxy.}%
\label{abund}%
\end{center}
\end{figure}

In figure \ref{abund} we show the relative distribution of elements
in our galaxy. It has two distinct regions: in the region $A<100$ it
decreases with $A$ approximately like an exponential, whereas for
$A>100$ it is approximately constant, except for the peaks in the
region of the magic numbers $Z=50$ and
$N=50,$ $82,$ 126.%

The thermonuclear processes (\ref{astroeq1})-(\ref{astroeq8}) can
explain the relative abundance of $_{2}^{4}$He, $_{\ 6}^{12}$C and
$_{\ 8}^{16}$O. The processes occurring after $_{2}^{4}$He burning
mainly form isotopes of $_{10}^{20}$Ne, $_{12}^{24}$Mg and
$_{14}^{28}$Si. We can understand the small abundance of the
elements Li, Be and B as due to the small velocity with which they
are formed via the reaction (\ref{astroeq4}) and the first equation
of (\ref{astroeq5}), while they are rapidly consumed by the second
reaction in (\ref{astroeq5}) and the first reaction in
(\ref{astroeq6}).

The synthesis of elements heavier than oxygen occur when, after
helium burning, a new compression and heating of the star rises the
temperature to values higher than $6\times10^{8}$ K. This situation
ignites an intense carbon burning:
\begin{eqnarray}
\ \ _{\ 6}^{12}\mathrm{C}+_{\ 6}^{12}\mathrm{C}\ \ \longrightarrow
\ \ _{10}^{20}\mathrm{Ne}+\ _{2}^{4}\mathrm{He}\longrightarrow\ \ _{11}%
^{23}\mathrm{Na}+\ \mathrm{p}\nonumber \\\longrightarrow\ \ _{12}^{23}\mathrm{Mg}%
+\ \mathrm{n}\longrightarrow\ \ _{12}^{24}\mathrm{Mg}+\ \gamma
.\label{astrophys24}%
\end{eqnarray}

Carbon and oxygen can also burn simultaneously:
\begin{eqnarray}
\ \ _{\ 6}^{12}\mathrm{C}+\ \ _{\ 8}^{16}\mathrm{O}\longrightarrow
\ \ _{12}^{24}\mathrm{Mg}+\ _{2}^{4}\mathrm{He}%
,\ \mathrm{etc,}\nonumber \\
 _{\ 8}^{16}\mathrm{O}+\ \ _{\ 8}%
^{16}\mathrm{O}\longrightarrow\ \ _{14}^{28}\mathrm{Si}+\ _{2}^{4}%
\mathrm{He},\ \mathrm{etc.}\label{astrophys25}%
\end{eqnarray}

For temperatures above $3\times10^{9}\ $K more photo-nuclear
processes appear. These yield more nuclei to be burned and heavier
nuclei are produced:
\begin{eqnarray}
\gamma+\ \ _{14}^{28}\mathrm{Si}\longrightarrow\ \ _{12}^{24}\mathrm{Mg}%
+\ _{2}^{4}\mathrm{He},\nonumber \\
 _{2}^{4}\mathrm{He}+\ \ _{14}^{28}%
\mathrm{O}\longrightarrow\ \ _{16}^{32}\mathrm{S}+\ \gamma,\ \mathrm{etc.}%
\label{astrophys26}%
\end{eqnarray}

Due to the large number of free neutrons, many
(n,$\gamma$)-reactions (radiative neutron capture) elements in the
mass range $A=28,\ldots,57$\ are formed. This leads to a large
abundance of elements in the iron mass region, which have the
largest binding energy per nucleon. For elements heavier than
iron the nuclear fusion processes do not generate energy.%

For $A>100$ the distribution of nuclei cannot be explained in terms
of fusion reactions with charged particles. They are formed by the
successive capture of slow neutrons and of $\beta^{-}$-decay. The
maxima of the element distribution in $N=50,$ $82,$ 126 are due to
the small capture cross sections corresponding to the magic numbers.
This yields a trash of isotopes at the observed element
distribution.

\subsection{Thermonuclear cross sections}

The nuclear cross section for a reaction between target $j$\ and
projectile
$k$\ is defined by%
\begin{equation}
\sigma=\frac{\mathrm{{number\ of\ reactions\ target^{-1}sec^{-1}}}%
}{\mathrm{flux\ of\ incoming\ projectiles}}={\frac{{r/n_{j}}}{{n_{k}v}}%
}.\label{astrophys1}%
\end{equation}
where the target number density is given by $n_{j}$, the projectile
number density is given by $n_{k}$\ , and v is the relative velocity
between target and projectile nuclei. Then $r$, the number of
reactions per cm$^{3}$\ and
sec, can be expressed as $r=\sigma vn_{j}n_{k}$, or, more generally,%
\begin{equation}
{\small r}_{j,k}{\small =}\int{\small \sigma|v}_{j}{\small -v}_{k}%
{\small |d}^{3}{\small n}_{j}{\small d}^{3}{\small n}_{k}{\small .}%
\label{astrophys2}%
\end{equation}

The evaluation of this integral depends on the type of particles and
distributions which are involved. For nuclei $j$\ and $k$\ in an
astrophysical
plasma, obeying a Maxwell-Boltzmann distribution (MB),%
\begin{equation}
d^{3}n_{j}=n_{j}({\frac{{m_{j}}}{{2\pi
kT}}})^{3/2}\mathrm{exp}(-{\frac
{{m_{j}v}^2_{j}}{{2kT}}})d^{3}{v}_{j},\label{astrophys3}%
\end{equation}
eq. (\ref{astrophys2}) simplifies to $r_{j,k}=<\sigma v>n_{j}n_{k}$,
where $<\sigma v>$\ is the average of $\sigma v$\ over the
temperature distribution in (\ref{astrophys3}).

In astrophysical plasmas with high densities and/or low
temperatures, effects of electron screening become highly important.
This means that the reacting nuclei, due to the background of
electrons and nuclei, feel a different Coulomb repulsion than in the
case of bare nuclei. Under most conditions (with non-vanishing
temperatures) the generalized reaction rate integral can be
separated into the traditional expression without screening and a
screening factor
\begin{equation}
<\sigma v>_{j,k}^{\ast}=f_{scr}(Z_{j},Z_{k},\rho,T,Y_{i})<\sigma
v>_{j,k}.
\end{equation}

This screening factor is dependent on the charge of the involved
particles,
the density, temperature, and the composition of the plasma. Here $Y_{i}%
$\ denotes the abundance of nucleus $i$\ defined by $Y_{i}=n_{i}/(\rho N_{A}%
)$, where $n_{i}$\ is the number density of nuclei per unit volume and $N_{A}%
$\ Avogadro's number. At high densities and low temperatures
screening factors can enhance reactions by many orders of magnitude
and lead to pycnonuclear ignition.

When in eq. (\ref{astrophys2}) particle $k$\ is a photon, the
relative velocity is always $c$\ and quantities in the integral are
not dependent on $d^{3}n_{j}$. Thus it simplifies to
$r_{j}=\lambda_{j,\gamma}n_{j}$\ and $\lambda_{j,\gamma}$\ results
from an integration of the photodisintegration cross section over a
Planck distribution for photons of temperature $T$
\begin{equation}
r_{j} =\frac{1}{{\pi^{2}(c\hbar)^{3}}}{{\int
d^{3}n_{j}}}\int_{0}^{\infty}{\frac{{c\sigma(E_{\gamma})E_{\gamma}^{2}}%
}{{\mathrm{exp}(E_{\gamma}/kT)-1}}}dE_{\gamma}.\label{astrophys7}%
\end{equation}

There is, however, no direct need to evaluate photodisintegration
cross sections, because, due to detailed balance, they can be
expressed by the capture cross sections for the inverse reaction
$l+m\rightarrow j+\gamma $.

A procedure similar to eq. (\ref{astrophys7}) is used for electron
captures by nuclei. Because the electron is about 2000 times less
massive than a nucleon, the velocity of the nucleus $j$\ is
negligible in the center of mass system in comparison to the
electron velocity ($|v_{j}-v_{e}|\approx|v_{e}|$). The electron
capture cross section has to be integrated over a Boltzmann,
partially degenerate, or Fermi distribution of electrons, depending
on the astrophysical conditions. The electron capture rates are a
function of $T$\ and $n_{e}=Y_{e}\rho N_{A}$, the electron number
density \cite{FFN85}. In a neutral, completely ionized plasma, the
electron abundance is equal to the total proton abundance in nuclei
$Y_{e}=\sum_{i}Z_{i}Y_{i}$\ and
${\small r}_{j}{\small =\lambda}_{j,e}{\small (T,\rho Y}_{e}{\small )n}%
_{j}{\small .} $

This treatment can be generalized for the capture of positrons,
which are in a thermal equilibrium with photons, electrons, and
nuclei. At high densities ($\rho>10^{12}$ g.cm$^{-3}$) the size of
the neutrino scattering cross section on nuclei and electrons
ensures that enough scattering events occur to thermalize a neutrino
distribution. Then also the inverse process to electron capture
(neutrino capture) can occur and the neutrino capture rate can be
expressed similarly to Eq. (\ref{astrophys7}), integrating over the
neutrino distribution. Also inelastic neutrino scattering on nuclei
can be expressed in this form. Finally, for normal decays, like beta
or alpha decays with half-life $\tau_{1/2}$, we obtain an equation
similar to Eq. (\ref{astrophys7}) or $r_j$ of the last paragraph
with a decay constant $\lambda_{j}=\ln2/\tau_{1/2}$\ and $
{r}_{j}=\lambda_{j}{n}_{j}$.

\section{Reactions with radioactive nuclear beams}

The basic research activity in nuclear physics, driven by the desire
to understand the forces which dictate the properties of nuclei, has
spawned a large number of beneficial applications. Amongst its many
progeny we can count reactor- and spallation-based neutron sources,
synchrotron radiation sources, particle physics, materials
modification by implantation, carbon dating and much more. It is an
excellent example of the return to society of investment in basic
research.

All of these achievements have been realized by accelerating the 283
stable or long-lived nuclear species we find here on Earth.  In
recent years, however, it has become evident that it is now
technically possible to create and accelerate unstable nuclei and,
and there are some 6-7,000 distinct nuclear species which live long
enough to be candidates for acceleration. They are the nuclei within
the so-called {\it drip-lines}, the point where the nucleus can no
longer hold another particle. This has led to many new opportunities
in industry, medicine, material studies and the environment.

\begin{figure}
[t]
\begin{center}
\includegraphics[
natheight=3.062300in, natwidth=10.312900in, height=1.in, width=3.2in
]%
{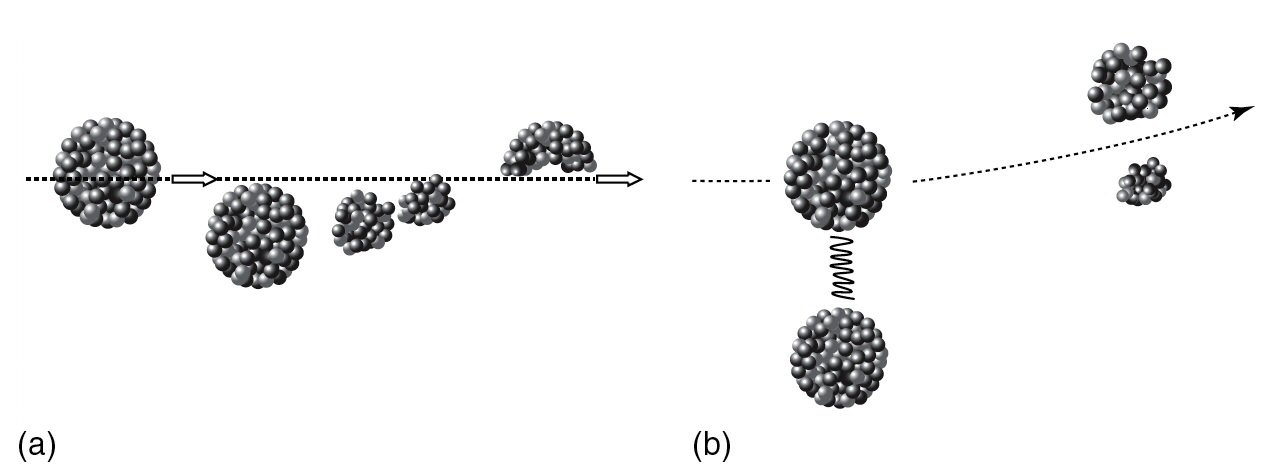}
\caption{(a) Schematic description of a nuclear fragmentation
reaction producing rare \ isotopes. The lower fragments are called
participants, while the upper one is called the spectator. Using
uranium projectiles ($N/Z\sim1.6$) one expects to produce (light)
spectator nuclei of about the same $N/Z$\ ratio. (b) Coulomb fission
of relativistic projectiles leading to the production of rare
isotopes. For a heavy unstable projectile an exchanged photon with
the target can give it enough energy to fragment into several
types of isotopes. }%
\label{f1}%
\end{center}
\end{figure}

Assume that a highly energetic uranium projectile $\,(N/Z\sim1.6)\,$
hits a target nucleus in an almost central collision, as shown in
fig. \ref{f1}. A part of the projectile ({\it participant}) is
scrapped off and forms a highly excited mixture of nucleons with a
part of the target. A piece of the projectile ({\it spectator})
flies away with nearly the same velocity of the beam. The
neutron-to-proton ratio of the spectator is nearly equal to that of
the projectile. Since the $\,N/Z$ - ratio of light nuclei \ (stable)
is close to one, the fragment is far from the stability line.
Statistically, a large number of fragments with different $\,N/Z$ -
ratios is created and several new exotic nuclei have been discovered
in this way. These nuclei can be collected in a secondary beam,
further accelerated and induce reactions with a target nucleus. This
method has become an important tool to study the properties of
short-lived isotopes.

Experiments with secondary-beams are limited by reaction cross
section and {\it luminosity}. The luminosity $L$ is defined as the
product
of beam intensity $i $ and target thickness $t$, $L=i\cdot t\label{luminosity}%
$. The reaction rate $N$ is the product of luminosity and reaction
cross section $\sigma_{r}$, $
N=\sigma_{r}\cdot L.\label{reactionrate}%
$. In most of the reactions the usable target thickness is limited
by the width of the excitation function (i.e. the cross section as a
function of the excitation energy). Production reactions with a wide
excitation function covering a broad energy range can profit in
luminosity by the use of thick targets.

The condition for fragmentation of heavy ions is that the projectile
should move faster than nucleons move inside the nucleus. The
projectile energy should be sufficiently above the Fermi domain,
e.g. above 100 A MeV. The usable target thickness for these high
energies is of the order of several grams per square centimeter,
corresponding to $10^{23}$ atoms/cm$^{2}$. The excitation function
for complete fusion of heavy ions, however, has a width of only 10
MeV. This corresponds to a usable target thickness of the order of
one milligram per square centimeter or $10^{18}$ atoms/cm$^{2}$.
Consequently beam intensities for the investigation of complete
fusion reactions must be by four
to five orders higher to achieve the same luminosity as for fragmentation.%

\begin{figure}
[t]
\begin{center}
\includegraphics[
natheight=14.437200in, natwidth=15.155800in, height=2.2in,
width=2.2in
]%
{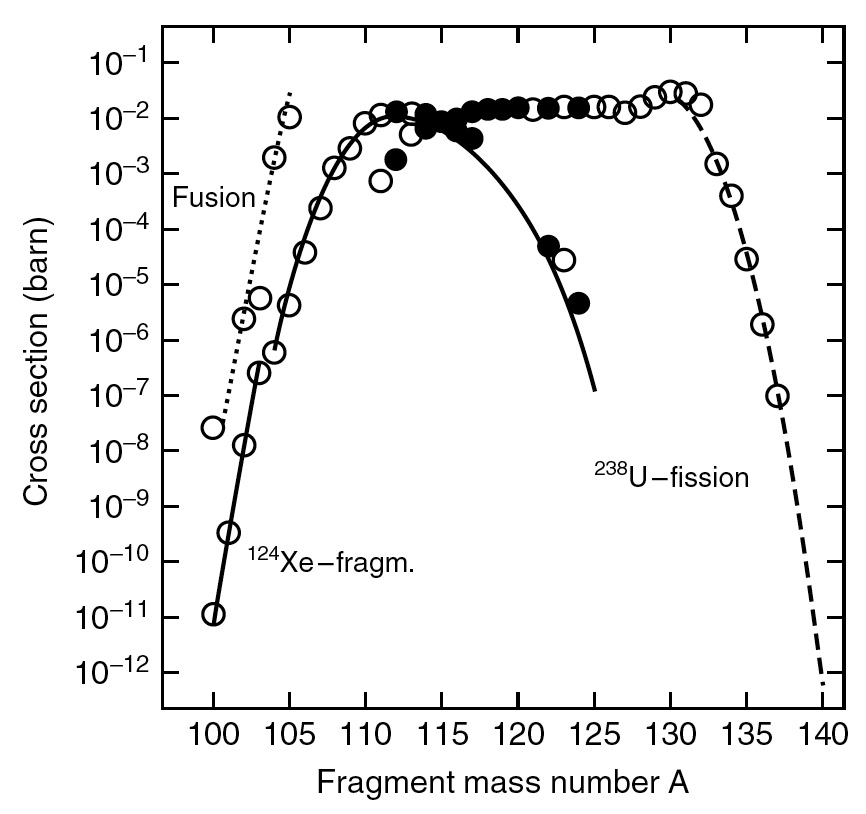}
\caption{\sl Production cross sections for the tin isotopes from
complete fusion (dotted line), fragmentation (solid line), and
projectile fission of $^{238}\mathrm{U}$ (dashed line). The symbols
represent experimental data. The fragmentation cross-sections (solid
line) have been calculated with a
semi-empirical code [Sue00].}%
\label{erice1}%
\end{center}
\end{figure}

Fig. \ref{erice1} shows as an example the production cross sections
for the tin isotopes from complete fusion (dotted line), nuclear
fragmentation (solid line), and Coulomb fission of
$^{238}\mathrm{U}$ (dashed line). The symbols represent experimental
data. The fragmentation cross-sections (solid line) have been
calculated with a semi-empirical code [Sue00].

The first experiments with unstable nuclear beams were designed to
measure the nuclear sizes, namely the matter distribution of protons
and neutrons. For stable nuclei such experiments are best
accomplished with electron beams, which probe the nuclear charge
(proton) distribution. Electron scattering experiments with unstable
beams can only be performed in an electron-nucleus collider.  But
the easiest solution is to measure the \textit{reaction cross
section} in collisions of unstable beams with a fixed target nucleus
\cite{Ta85}.

The reaction cross section in high energy collisions is given by
\begin{equation}
\sigma_{R}=2\pi\int[1-T(b)]bdb\;,\label{sigreac}%
\end{equation}
where
\begin{equation}
T(b)=\exp\left\{  -\sigma_{NN}\,\int_{-\infty}^{\infty}dz\int\,\rho
_{P}\,(\mathbf{r}\,)\rho_{T}\left(  \mathbf{R}+\mathbf{r}\right)
d^{3}r\right\}  \label{tb}%
\end{equation}
with $\,\mathbf{R}=(\mathbf{b},z)\,$. $\sigma_{NN}\,$ is the
nucleon-nucleon cross section at the corresponding bombarding
energy, and $\,\rho_{P(T)}\,$ is the projectile (target) matter
density distribution. $\,T(b)\,$ is known as the
\textit{transparency function}. It is the probability that a
reaction occurs for a collision with  impact parameter $b$. The
exponent $\left[  \sigma
_{NN}\,\int\,\rho_{P}\,(\mathbf{r}\,)\rho_{T}\left(  \mathbf{R}+\mathbf{r}%
\right)  d^{3}r\right]  ^{-1}$ is interpreted as the \ mean-free
path for a nucleon-nucleon collision. The reaction cross sections
can be calculated using the matter distributions of the target in
eq.~\ref{sigr}.

\begin{figure}
[t]
\begin{center}
\includegraphics[
natheight=6.198100in, natwidth=9.781000in, height=2.8in, width=2.6in
]%
{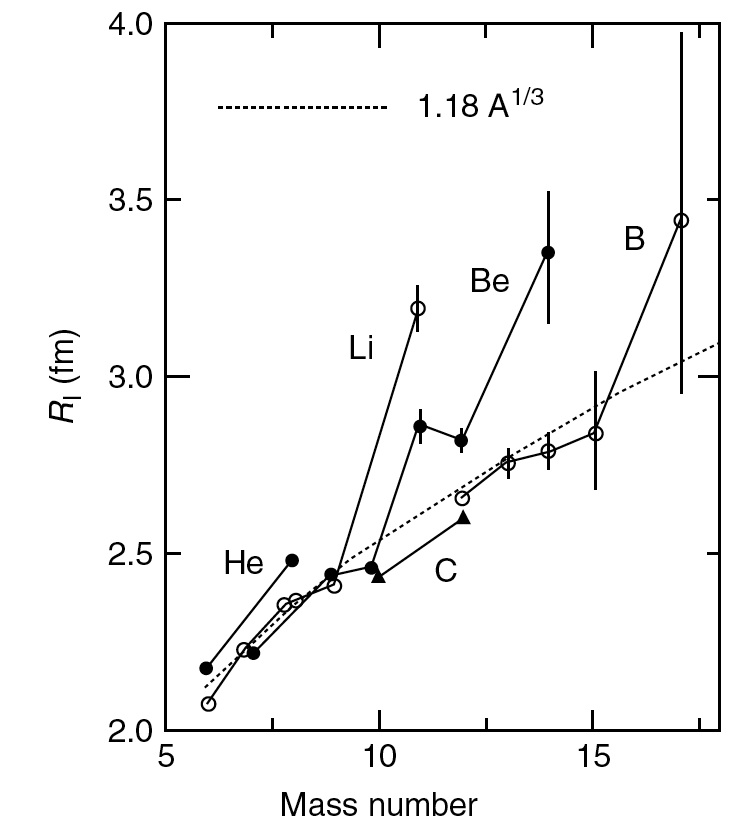}
\caption{The matter density radii of several light nuclei compared
to the trend $\,R\sim1.18\;\;A^{1/3}$ fm \ (dashed line) for normal
nuclei. The solid lines
are guides to the eyes.}%
\label{sizes}%
\end{center}
\end{figure}

Experimentally it was observed a great increase in the rms radii for
the neutron-rich isotopes $\,^{6}$He, $\,^{8}$He and $^{11}$Li (see
figure \ref{sizes}). Thus, the addition of the neutrons to $^{4}$He
and $^{9}$Li nuclei increase their radii considerably. This can be
understood in terms of the binding energy of the outer nucleons. The
large matter radii of these nuclei have lead the experimentalists to
call them \textquotedblleft\emph{halo nuclei}\textquotedblright. The
binding energy of the last two neutrons in $^{11}$Li is about
$\,300$~keV. In $^{6}$He it is
0.97~MeV. These are very small values and should be compared with $S_{n}%
=6-8$~MeV which is the average binding of nucleons in stable nuclei.
Abnormally large radii were also found for other light neutron-rich
nuclei.

The matter density radii of these nuclei do not follow the observed
trend $\,R\sim1.18\;\;A^{1/3}$ fm \ of normal nuclei. Thus the halo
seems to be a common feature of loosely-bound neutron-rich nuclei.

Several other methods have been devised to probe the structure of
nuclei far from the valley of stability. Among these are (a) Coulomb
dissociation \cite{BBR86}, (b) Trojan horse method \cite{Bau86}, (c)
asymptotic normalization coefficients  \cite{Xu04}, (d) heavy ion
charge-exchange \cite{St96}, (e) knockout \cite{HT03} and (f) fusion
reactions \cite{Can06}. These methods yield different insights into
the structure of exotic nuclei and comprise most of experiments in
radioactive beam facilities \cite{Gee06}.

Reactions producing rare nuclear isotopes has opened a new research
front in nuclear physics with applications in many areas of science:
(a) the rapid production of short-lived nuclei, many of which are
poorly known, is important for astrophysics (r-process) and
cosmology, (b) nuclear medicine benefits from studies of new nuclear
isotopes, (c) and the list goes on. But the basic question still
remains: what combinations of neutrons and protons can make up a
nucleus? The experimental detection of new nuclear isotopes is an
ongoing research which will ultimately lead to new insight and
development of nuclear science, with enormous profit for mankind
\cite{Bau07}.

\acknowledgements The author is grateful to Prof. Reinhard Stock and
Prof. Konrad Gelbke for their support and encouragement. This work
was supported in part by funds provided by the U.S. Department of
Energy (DOE) under contract No. DE-FG02-08ER41533 and DE-FC02-07ER41457
(UNEDF, SciDAC-2), and the Research Corporation under
Award No. 10497.

\bigskip
{\bf Further Reading}

C.A. Bertulani and P. Danielewicz, {\it Introduction to Nuclear
Reactions}, IOP Publishing,  London, 2004.

C.A. Bertulani, {\it Nuclear Physics in a Nutshell}, Princeton
University Press, Princeton, 2007.

Pawel Danielewicz, Roy Lacey and William G. Lynch, {\it
Determination of the Equation of State  of Dense Matter}, Science
298, 1592 (2002).

Herman Feshbach, {\it Theoretical Nuclear Physics, Nuclear
Reactions}, Wiley-Interscience, New York, 1993.

P. Fr\"obrich and R. Lipperheide, {\it Theory of Nuclear Reactions},
Oxford University Press, Oxford, 1996.

Christian Iliadis, {\it Nuclear Physics of Stars},  Wiley, New York,
(2007).

J.S. Lilley, {\it Nuclear Physics: Principles and Applications},
Wiley, New York, 2001.

Norman K. Glendenning, {\it Direct Nuclear Reactions}, World
Scientific, Singapore, 2004.

Samuel S. M.  Wong,  {\it Nuclear Physics},
 Wiley, New York, 2004.

C.Y. Wong,  {\it Introduction to high energy heavy ion collisions},
World Scientific, Singapore, 1994.

\end{document}